\documentclass[11pt,titlepage]{article} 
\usepackage{amsfonts}
\usepackage{amsmath}
\usepackage{amssymb}
\usepackage{tipa}
\usepackage{url}
\usepackage{pdfpages}
\usepackage[square,numbers]{natbib}

\usepackage{caption}
\usepackage{subfigure}

\usepackage{setspace}

\begin{document}
\title{Ancient west Eurasian ancestry in southern and eastern Africa}
 \small \author{Joseph K. Pickrell$^{1, \dagger}$, Nick Patterson$^2$, Po-Ru Loh$^3$, Mark Lipson$^3$, Bonnie Berger$^{2,3}$,\\
  Mark Stoneking$^4$,  Brigitte Pakendorf$^{5, \dagger}$, David Reich$^{1,2,\dagger}$\\
  \\
  \small $^1$ Department of Genetics, Harvard Medical School\\
  \small $^2$ Broad Institute \\
\small $^3$ Department of Mathematics and Computer Science and AI Laboratory, MIT\\
  \small $^4$ Department of Evolutionary Genetics, MPI for Evolutionary Anthropology\\
    \small $^5$ Laboratoire Dynamique du Langage, UMR5596, CNRS and Universit\`e Lyon Lumi\'ere 2 \\
    \small $^\dagger$ To whom correspondence should be addressed: \url{joseph_pickrell@hms.harvard.edu}, \\
    \small \url{brigitte.pakendorf@cnrs.fr},  \url{reich@genetics.med.harvard.edu}\\
  }
\maketitle
\pagebreak

\begin{abstract}
The history of southern Africa involved interactions between indigenous hunter-gatherers and a range of populations that moved into the region. Here we use genome-wide genetic data to show that there are at least two admixture events in the history of Khoisan populations (southern African hunter-gatherers and pastoralists who speak non-Bantu languages with click consonants). One involved populations related to Niger-Congo-speaking African populations, and the other introduced ancestry most closely related to west Eurasian (European or Middle Eastern) populations. We date this latter admixture event to approximately 900-1,800 years ago, and show that it had the largest demographic impact in Khoisan populations that speak Khoe-Kwadi languages. A similar signal of west Eurasian ancestry is present throughout eastern Africa. In particular, we also find evidence for two admixture events in the history of Kenyan, Tanzanian, and Ethiopian populations, the earlier of which involved populations related to west Eurasians and which we date to approximately 2,700 - 3,300 years ago. We reconstruct the allele frequencies of the putative west Eurasian population in eastern Africa, and show that this population is a good proxy for the west Eurasian ancestry in southern Africa. The most parsimonious explanation for these findings is that west Eurasian ancestry entered southern Africa indirectly through eastern Africa.
\end{abstract}

\section*{Introduction}
Hunter-gatherer populations have inhabited southern Africa for tens of thousands of years \citep{phillipson2005african}. Within approximately the last two thousand years, these populations were joined by food-producing groups (both pastoralists and agriculturalists), and a culturally diverse set of populations occupy the region today. Because written history is unavailable until recently in southern Africa, inferences about the migration patterns leading to the present distribution of populations have largely been informed by archaeology and linguistics.

Genetic data is an additional source of information about population history, but extracting this information remains challenging. Studies of diversity in southern Africa have highlighted the influence of pre-colonial population admixture on the genetic structure of populations in the region \citep{Schlebusch:2012uq, Pickrell:2012fk, Petersen:2013fk}, but have come to different conclusions about the historical scenarios that led to this admixture. In particular, though there is agreement that the arrival of Bantu-speaking agriculturalist populations had a major demographic impact in many populations, the importance of population movements from other parts of Africa or the world is unclear. Schlebusch et al. \citep{Schlebusch:2012uq} argued for eastern African ancestry specifically in the Nama, a pastoralist population, while Pickrell et al. \citep{Pickrell:2012fk} raised this possibility not just for the Nama, but for several Khoe-speaking populations. Identifying the sources of non-Khoisan ancestry in southern Africa could shed light on the historical processes that led to the extensive linguistic and cultural diversity of the region. 

Here, we use new techniques based on the extent of linkage disequilibrium to thoroughly examine the signal of admixture in the southern African Khoisan (defined here as indigenous populations speaking non-Bantu languages with click consonants, without implying cultural, linguistic, or genetic homogeneity of Khoisan groups). First, we show that all Khoisan populations have some non-zero proportion of west Eurasian ancestry. (Throughout this paper, we will use geographic labels to refer to ancestry, with the caveat that the geographic labels are derived from modern populations--that is, when we refer to ``west Eurasian ancestry" in ``southern Africa", we are using this as a shorthand for the more cumbersome, but more accurate, phrasing of ``ancestry most closely related to populations currently living in west Eurasia" in ``populations currently living in southern Africa".) Second, we show that there are multiple waves of population mixture in the history of many southern and eastern African populations, and that west Eurasian ancestry entered eastern Africa on average 2,700-3,300 years ago and southern Africa 900-1,800 years ago. Third, we infer the allele frequencies of the ancestral west Eurasian population in eastern Africa, and show that this population is a good proxy for the west Eurasian ancestry in southern Africa. We thus argue that the most plausible source of west Eurasian ancestry in southern Africa is indirect gene flow via eastern Africa.
\section*{Results}
We began with an analysis of population mixture in southern Africa, using the data from Pickrell et al. \citep{Pickrell:2012fk} supplemented with an additional 32 individuals from seven Khoisan populations genotyped on the Affymetrix Human Origins Array (Supplementary Table 1); note that the Damara are excluded from most of the subsequent analyses as they genetically resemble southern African Bantu-speaking groups \citep{Pickrell:2012fk}. These southern African data were then combined with previously-published worldwide data \citep{Patterson:2012fk} (SI Methods). After removing individuals who appeared to be genetic outliers with respect to others in their population (SI Methods), we analyzed a final data set consisting of 1,040 individuals from 75 worldwide populations, all genotyped on the Affymetrix Human Origins array at 565,259 single nucleotide polymorphisms (SNPs). These data are available on request from the authors for use in analyses of population history.

\subsection*{West Eurasian ancestry in the Ju$|$'hoan\_North.}
We previously observed that the Ju$|$'hoan\_North, though the least admixed of all Khoisan populations, show a clear signal of admixture when using a test based on the decay of admixture linkage disequilibrium (LD) \citep{Pickrell:2012fk}. The theoretical and practical aspects of historical inference from admixture LD have since been examined in greater detail \citep{alder}; we thus re-evaluated this signal in the Ju$|$'hoan\_North using the software ALDER v1.0 \citep{alder}. 

\begin{figure}
\begin{center}
\includegraphics[scale = 1]{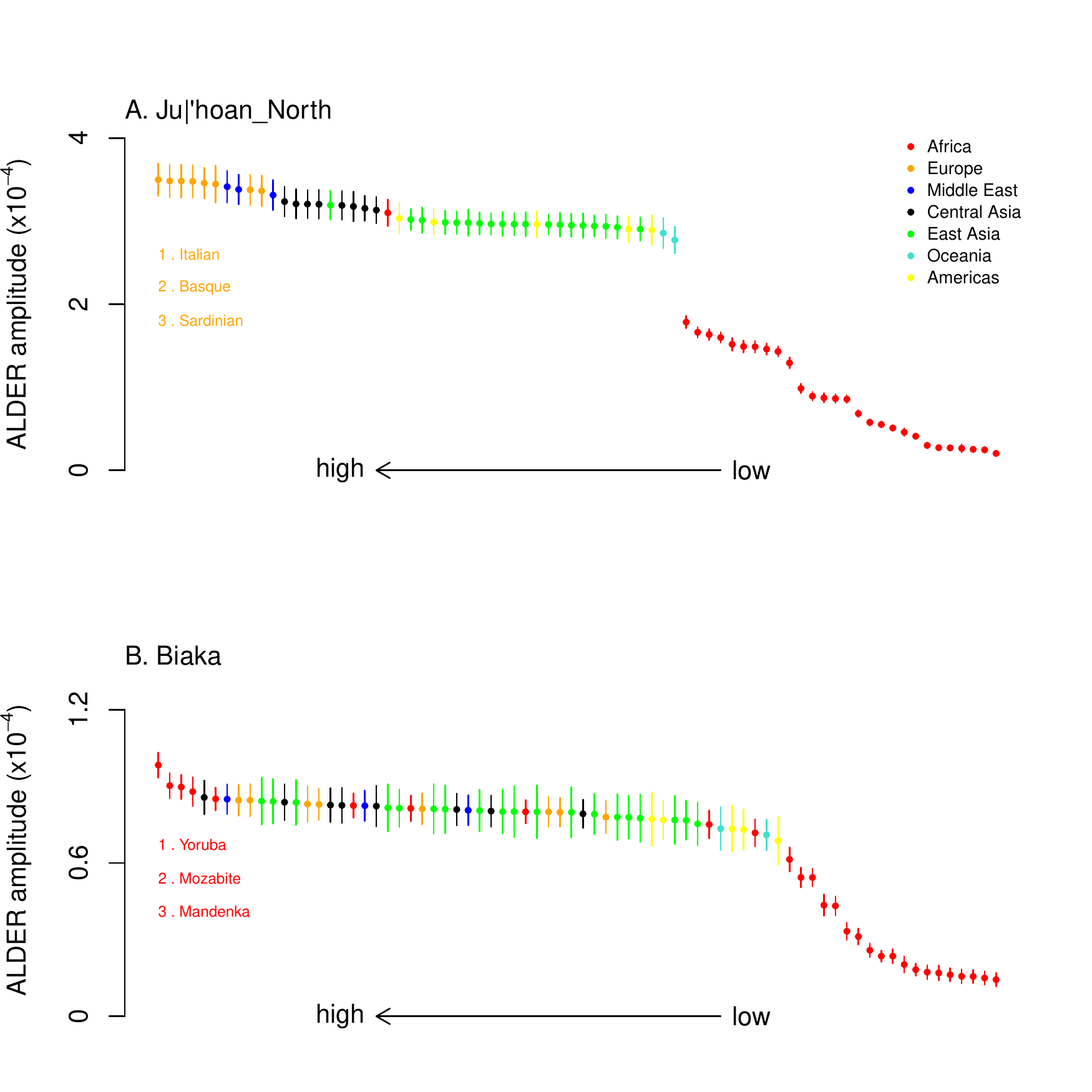}
\caption{\textbf{: Identifying the sources of admixture in the A) Ju$|$'hoan\_North and B) Biaka.} We computed weighted LD curves in  the Ju$|$'hoan\_North and Biaka using the test population itself as one reference and a range of other populations as the second reference. We then fitted an exponential decay curve to each LD curve, starting from 0.5 cM. Plotted are the fitted amplitudes for each curve. Error bars indicate one standard error. A larger amplitude indicates a closer relationship to one of the true admixing populations. Populations are ordered according to the amplitude, and colored according to their continent of origin. The three populations with the largest amplitude (and thus the closest inferred relationship to the true mixing population) are listed. Note that the only populations from western Africa in these data are the Yoruba and Mandenka.}\label{fig_ju_alder}
\end{center}
\end{figure}

In particular, we were interested in identifying the source of the gene flow by comparing weighted LD curves computed using different reference populations. This is possible because theory predicts that the amplitude of these curves (i.e., the average level of weighted LD between sites separated by 0.5 centimorgans) becomes larger as one uses reference populations that are closer to the true mixing populations. Loh et al. \citep{alder} additionally showed that this theory holds when using the admixed population itself as one of the reference populations. We thus computed weighted LD curves in the Ju$|$'hoan\_North, using the Ju$|$'hoan\_North themselves as one reference population and a range of 74 worldwide populations as the other, and examined the amplitudes of these curves (Figure \ref{fig_ju_alder}A). The largest amplitudes are obtained with European populations as references (Figure \ref{fig_ju_alder}A); taken literally, this would seem to implicate Europe as the source of admixture. The estimated date for this gene flow is 43 $\pm$ 2 generations (1290 $\pm$ 60 years, assuming 30 years/generation \citep{fenner2005cross}) before the present, consistent with our previously estimated date \citep{Pickrell:2012fk}. This date is well before the historical arrival of European colonists to the region. 

We next tested the robustness of this result. We confirmed that this observation is consistent across panels of SNPs with varied ascertainment (Supplementary Figure 2). We then considered hunter-gatherer populations from other regions of Africa. In particular, we performed the same analysis on the Biaka (Figure \ref{fig_ju_alder}B) and Mbuti (Supplementary Figure 3) from central Africa. As expected, the inferred source of admixture in these populations is a sub-Saharan African population (most closely related to the Yoruba, a Niger-Congo-speaking agriculturalist group from Nigeria).

A signal of west Eurasian ancestry in the Ju$|$'hoan\_North should be identifiable by allele frequencies as well as by LD. We thus tested the population tree [Chimp,[Ju$|$'hoan\_North, [Han, French]] using an $f_4$ statistic \citep{Reich:2009fk, Keinan:2007kx}. This tree fails with a Z-score of 4.0 ($P  = 3 \times 10^{-5}$). On smaller subsets of SNPs, the evidence is weaker, explaining why we had not noticed it previously (on the set of SNPs ascertained in a Ju$|$'hoan individual, Z = 2.7 [$P = 0.003$]; in a French individual, Z = 0.6 [$P = 0.27$]; in a Yoruba individual, Z = 1.4 [$P = 0.08$]). We thus conclude that there is a signal in both allele frequencies and linkage disequilibrium that the Ju$|$'hoan\_North admixed with a population more closely related to western rather than eastern Eurasian populations, and that this signal is absent from hunter-gatherer populations in central Africa.

\subsection*{Signal of west Eurasian relatedness is shared throughout southern Africa.}
\begin{figure}
\begin{center}
\includegraphics[scale = 1]{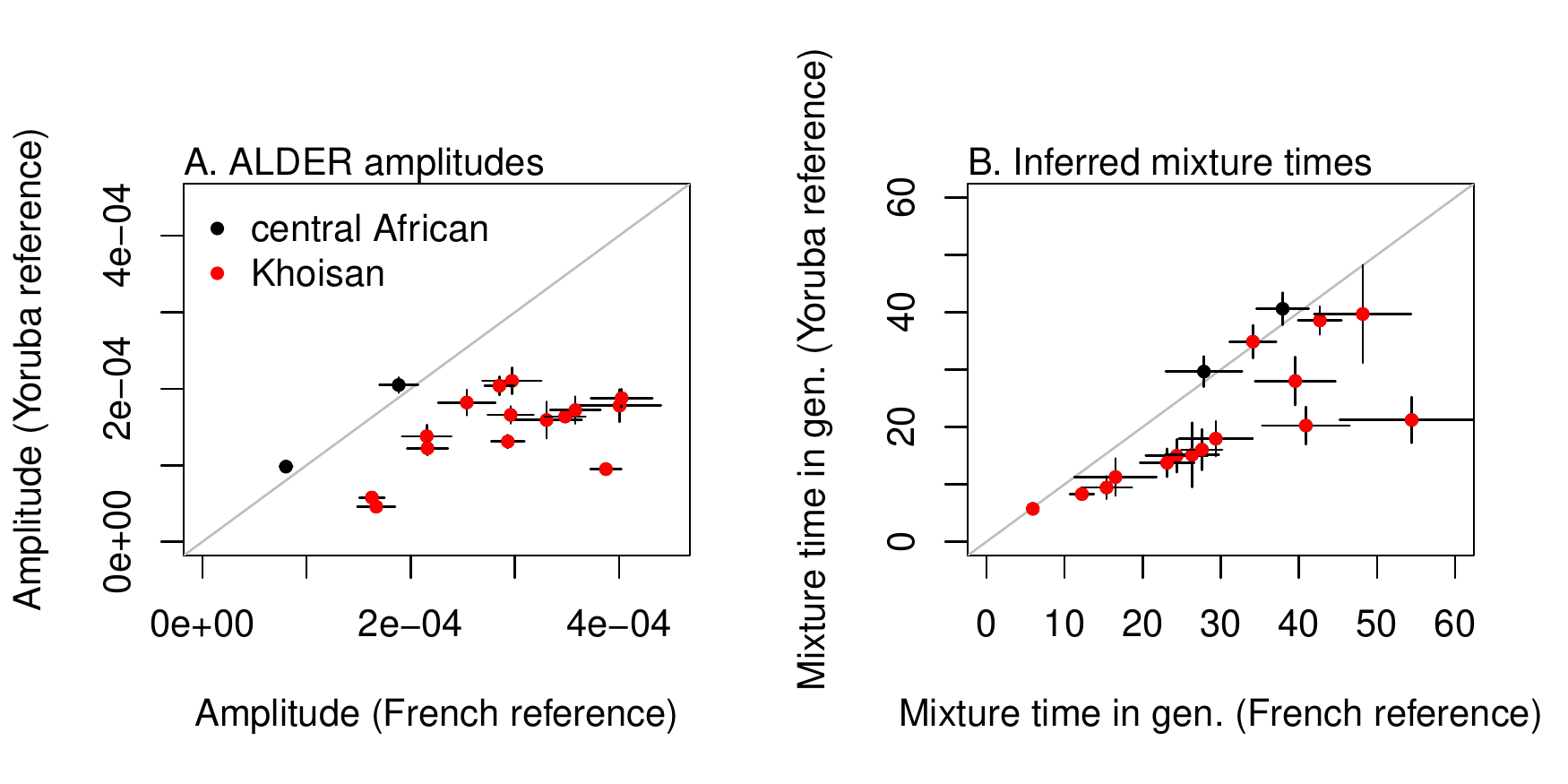}
\caption{\textbf{: Relationship with west Eurasia is shared by all Khoisan populations.} We generated weighted LD decay curves in each Khoisan (or central African hunter-gatherer) population, using weights computed using the test population as one reference and either the French or the Yoruba as the other reference. We then fit an exponential decay model to each LD curve. Plotted are the inferred \textbf{A)} amplitudes and \textbf{B)} admixture times in each population. Larger amplitudes indicate a closer relationship to the true admixing population, and under a model of a single admixture event, the admixture times do not depend on the reference populations used.}\label{fig_other_alder}
\end{center}
\end{figure}

We next examined whether this signal of relatedness to west Eurasia is present in other Khoisan populations. For each Khoisan population, we used ALDER to compute weighted LD decay curves using the test population as one reference and either the French or the Yoruba as the other reference. We included the central African Mbuti and Biaka populations as negative controls. In all Khoisan populations, the amplitude of the LD decay curve is larger when using the French as a reference than when using the Yoruba as a reference (Figure \ref{fig_other_alder}A). In contrast, for the Mbuti and Biaka, the larger amplitude is seen when using the Yoruba as a reference (Figure \ref{fig_other_alder}A). 

A striking observation that emerges from this analysis is that, in many of the southern African populations, the inferred mixture times depend substantially on the second population used as a reference (Figure \ref{fig_other_alder}B). Under a model of admixture from a single source population, the decay rate of the LD curve does not depend on the reference population used \citep{alder}; this suggests that there are at least two separate non-Khoisan sources of ancestry in some of these Khoisan populations. In contrast, for the central African Mbuti and Biaka, the inferred times do not depend on the reference used.

\subsection*{Estimating parameters of multiple admixture events.}
Motivated by the above observations, we designed a method to estimate dates of multiple admixture events in the history of a population (related ideas have been explored by Myers et al. \citep{Myers:2011z}). We extended the population genetic theory of Loh et al. \citep{alder} to the case where a population has experienced multiple episodes of population admixture from different sources (SI Methods). In this situation, the extent of admixture LD in the population is no longer a single exponential curve as a function of genetic distance, but instead a \emph{mixture} of exponential curves. Using a range of reference populations, we can thus formally test for the presence of multiple waves of mixture and estimate the dates of these mixture events (SI Methods). We validated this approach using coalescent simulations of three pairs of mixture dates chosen to span the scenarios that our data suggest are relevant to southern and eastern Africa (SI Methods). The simulations indicate that our method has reasonable but not perfect power; depending on the pair of dates of we simulated, we successfully detected both events in between 70-90

\begin{figure}
\begin{center}
\includegraphics[scale = 1]{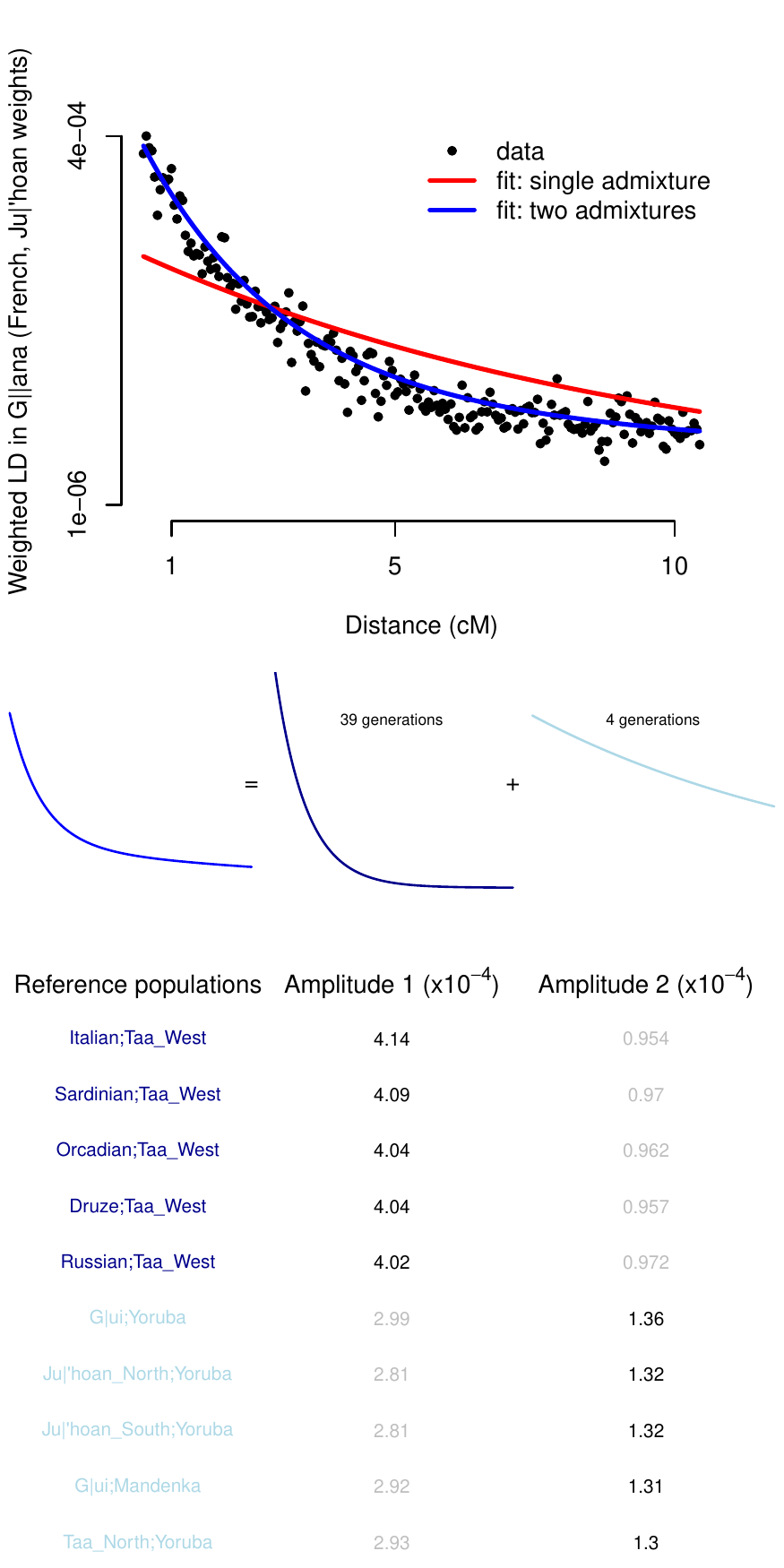}
\caption{\textbf{: LD evidence for multiple waves of mixture in the G$||$ana.} We computed 990 (45 choose 2) weighted LD curves in the G$||$ana, and fit two models: one with a single admixture event, and one with two admixture events. Shown is the LD curve computed using the French and Ju$|$'hoan\_North populations as references, along with the fitted curves from the two models (note that the decay rates in the fitted curves are shared across the data for all 990 pairs of populations, not only to the shown data). Below the plot, we show a schematic representation of the fitted model with two admixture events. In the table, we show the population pairs with the five largest estimated amplitudes on each admixture event (that is, the population pairs in dark blue are those with the largest weights on the dark blue curve, and those labeled in light blue are those with the largest weights on the light blue curve).}\label{fig_malder}
\end{center}
\end{figure}

To illustrate the intuition behind this method, in Figure \ref{fig_malder} we plot one of the weighted LD curves calculated in the G$||$ana. Under a model with a single admixture event, the mean admixture date in the G$||$ana is estimated as $14 \pm 3$ generations, identical to the date obtained by Pickrell et al. \citep{Pickrell:2012fk}. However, it is visually apparent that this model is a poor fit to the data (Figure \ref{fig_malder}). Indeed, we find that adding a second mixture event significantly improves the fit (minimum Z-score on the two admixture times of 2.8; $P = 0.003$) The two inferred mean admixture times in the G$||$ana are $4 \pm 1$ and $39\pm 6$ generations ago. 

This method additionally estimates amplitudes of the LD decay curves for each pair of populations on each mixture time, which are a function of the relationship between the reference populations and the true source populations. These amplitudes can be used to infer the references closest to the true mixing populations. However, if a source population is itself admixed, under some conditions this method will identify a population related to one of the ancestral components of the source population instead of the source population itself (SI Methods). By examining these amplitudes, we conclude that the west Eurasian ancestry in the G$||$ana entered the population through the older admixture event (Figure \ref{fig_malder}). Because of the caveat noted above, however, we cannot distinguish between two historical scenarios with this method: direct gene flow from a west Eurasian population and gene flow from a west Eurasian-admixed population. 
 
\begin{figure}
\begin{center}
\includegraphics[scale = 0.8]{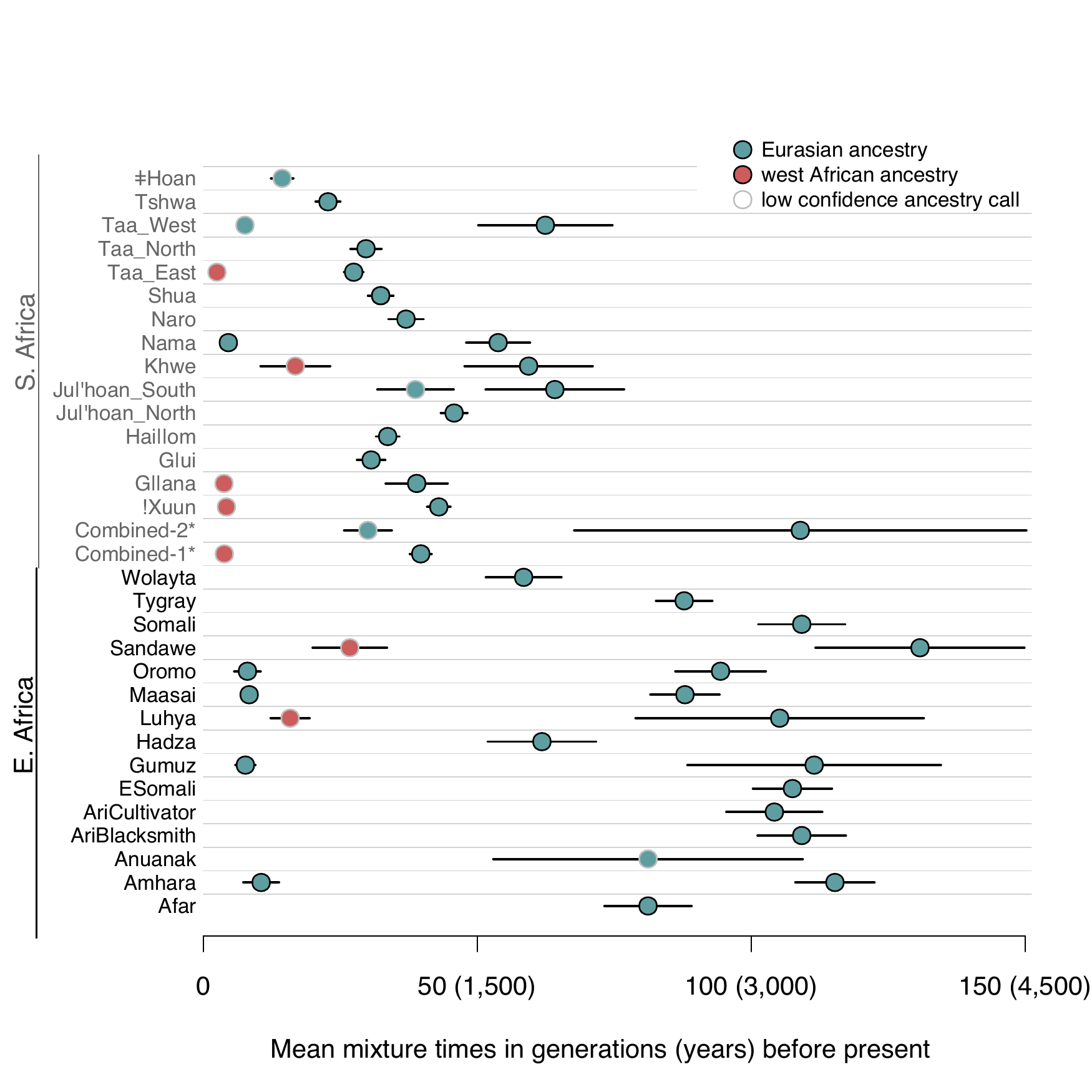}
\caption{\textbf{: Mean times of admixture in southern and eastern Africa.} For each southern or eastern African population, we estimated the number of mixture events and their dates. Plotted are the estimated dates. Black lines show one standard error on the estimates. Points are colored according to the populations inferred as proxies for the mixing populations (Methods). *The Combined-1 population is the Tshwa, Shua, Hai$||$om, \textdoublebarpipe Hoan, Naro, and Taa\_North. The Combined-2 population is the Ju$|$'hoan\_North and G$|$ui; see text for details. }\label{fig_illumina}
\end{center}
\end{figure}

We applied this method to each Khoisan population  in turn
 (with the exception of the Damara, who are genetically similar to non-Khoisan populations), using 45 other African and non-African populations as references(Supplementary Figures 10-23). In several populations, there is evidence for two waves of population mixture (!Xuun, Taa\_West, Taa\_East, Nama, Khwe, G$||$ana, Ju$|$'hoan\_South), while in others a single wave of population mixture fits the data (Figure \ref{fig_illumina}). For populations with two waves of mixture, west Eurasian ancestry entered through the earlier admixture event (Figure \ref{fig_illumina}, Supplementary Table 3). In the Nama, both the early and more recent admixture events are predicted to involve populations with west Eurasian ancestry, consistent with known post-colonial European admixture in this population. The Taa\_West also show two episodes of west Eurasian admixture, but the more recent one has low confidence. 

It is important to mention a few caveats in interpreting results from this method. First, in cases where the method detects two admixture events from the same source (as in the Nama and Taa\_West above), simulations suggest an alternative interpretation is sustained population mixture over many generations (SI Methods). Second, the numbers of admixture events inferred by this method are lower bounds; for example, this method fails to detect that the Naro are admixed between two distinct Khoisan groups \citep{Pickrell:2012fk}, and we find evidence of west African ancestry in just four Khoisan populations (!Xuun, G$||$ana, Khwe, and Taa\_East) when treated individually (but see analyses of combined populations below). Finally, the method has low confidence when assigning an admixture event to a population with west African ancestry (Figure 4); this reflects a relative lack of genetic drift specific to the west African reference populations (Yoruba and Mandenka), which makes it difficult to detect with high confidence (in contrast, there is considerable genetic drift in west Eurasian populations because of the out-of-Africa bottleneck, which allows admixture events to be more confidently assigned to this ancestry). 

In most populations where our method detects only a single admixture event, the fitted model visually appears inadequate to fully explain the data (e.g. Supplementary Figures 11, 13, 20, 21). Indeed, there is marginal statistical evidence for two admixture events in many of these populations (Supplementary Table 4). To increase our power to detect additional admixture events, we performed analyses of combined populations. In a combined set of populations (the Tshwa, Shua, Hai$||$om, \textdoublebarpipe Hoan, Naro, and Taa\_North) that have marginal evidence for a second, more recent admixture event, we infer two dates of admixture: one 40 $\pm$ 2 generations ago and one 4 $\pm$ 1 generations ago (Z-score for the hypothesis that the admixture time is zero is 3.2, $P=7\times 10^{-4}$). In a combined set of two populations (the Ju$|$'hoan\_North and G$|$ui) that have marginal evidence for a second, more ancient admixture event, we also infer two dates of admixture (Supplementary Figure 24), but with different dates from all other samples: one 30 $\pm$ 4 generations ago  (Z-score of 6.9, $P=2\times 10^{-12}$), and one 109 $\pm$ 41 generations ago (Z-score of 2.6, $P = 0.005$). We interpret this as suggestive evidence that the population that introduced west Eurasian ancestry to southern Africa was itself admixed, and that this more ancient admixture happened around 110 generations ago (though the confidence intervals here are clearly large). 

\subsection*{Variation in west Eurasian ancestry proportions in the Khoisan.}
We next asked if there are systematic differences between Khoisan populations in their levels of west Eurasian ancestry. To test this, we constructed an $f_4$ ratio estimate to specifically measure west Eurasian ancestry. This ratio is: $f_4$(Han, Orcadian; X, Druze)/ $f_4$(Han, Orcadian; Yoruba, Druze), where X is any southern African population; this ratio takes advantage of the fact that the west Eurasian ancestry is more closely related to Middle Eastern than to northern European populations (SI Methods). We applied this method to all Khoisan populations, and included southern African Bantu speakers for comparison. The highest levels of west Eurasian ancestry are found in Khoe-Kwadi speakers (Table \ref{frac_table}A), particularly the Nama, where our estimate of west Eurasian ancestry reaches 14\% (though note we cannot distinguish between the impact of recent colonialism and older west Eurasian ancestry in the Nama using this method). Other populations of note include the Khwe, Shua, and Hai$||$om, who we estimate to have approximately 5\% west Eurasian ancestry. The apparent correlation between language group and west Eurasian ancestry may have implications for the origins of this ancestry in southern Africa; we return to this point in the discussion.

\subsection*{The origin of west Eurasian ancestry in southern Africa.}
We next considered the origin of the west Eurasian ancestry in southern Africa. Direct interactions between Europe and southern Africa seem unlikely given the inferred admixture dates, especially because this ancestry is widespread throughout southern Africa. It has been reported that many populations in eastern Africa admixed with populations from the Levant \citep{Pagani:2012uq} or the Arabian peninsula \citep{Haber:2013fk}. Since there is suggestive genetic evidence of a migration from eastern Africa to southern Africa \citep{henn2008chromosomal, Pickrell:2012fk, Schlebusch:2012uq} as well as linguistic and archaeological indications  \citep{gldemann2008linguist},  we hypothesized that indirect gene flow through eastern Africa might be a plausible source for the west Eurasian ancestry in southern Africa. This hypothesis makes two major predictions: first, that the west Eurasian ancestry in eastern Africa should have the same source as that in southern Africa, and second, that the mixture times in eastern Africa should be older than those in southern Africa, perhaps with a date of around 110 generations (corresponding to the oldest date identified in southern Africa).

To test these predictions, we assembled a data set of individuals from southern Africa, eastern Africa, and west Eurasia typed on an Illumina platform by merging data from previous studies \citep{Pagani:2012uq, altshuler2010integrating, Behar:2010fk, Henn:2011fk, Li:2008rt}. The eastern African populations in these combined data include populations from Ethiopia, Kenya, Tanzania, and Sudan (the majority of these populations were genotyped by Pagani et al. \citep{Pagani:2012uq}). We first confirmed using $f_3$ tests \citep{Reich:2009fk} that many eastern African populations have statistically significant evidence for admixture with west Eurasian populations (Supplementary Table 5). The smallest $f_3$ statistics in nearly all eastern African populations involve a southern European (or Levantine) population as one reference. We then evaluated the fraction of west Eurasian ancestry in each population, using the same $f_4$ ratio estimate as used in the Khoisan (Table \ref{frac_table}B). The fraction of west Eurasian ancestry in eastern African populations is generally higher in eastern than in southern Africa; the highest levels of admixture (40-50\%) are observed in some Ethiopian populations.

\begin{figure}
\begin{center}
\includegraphics[scale = 0.8]{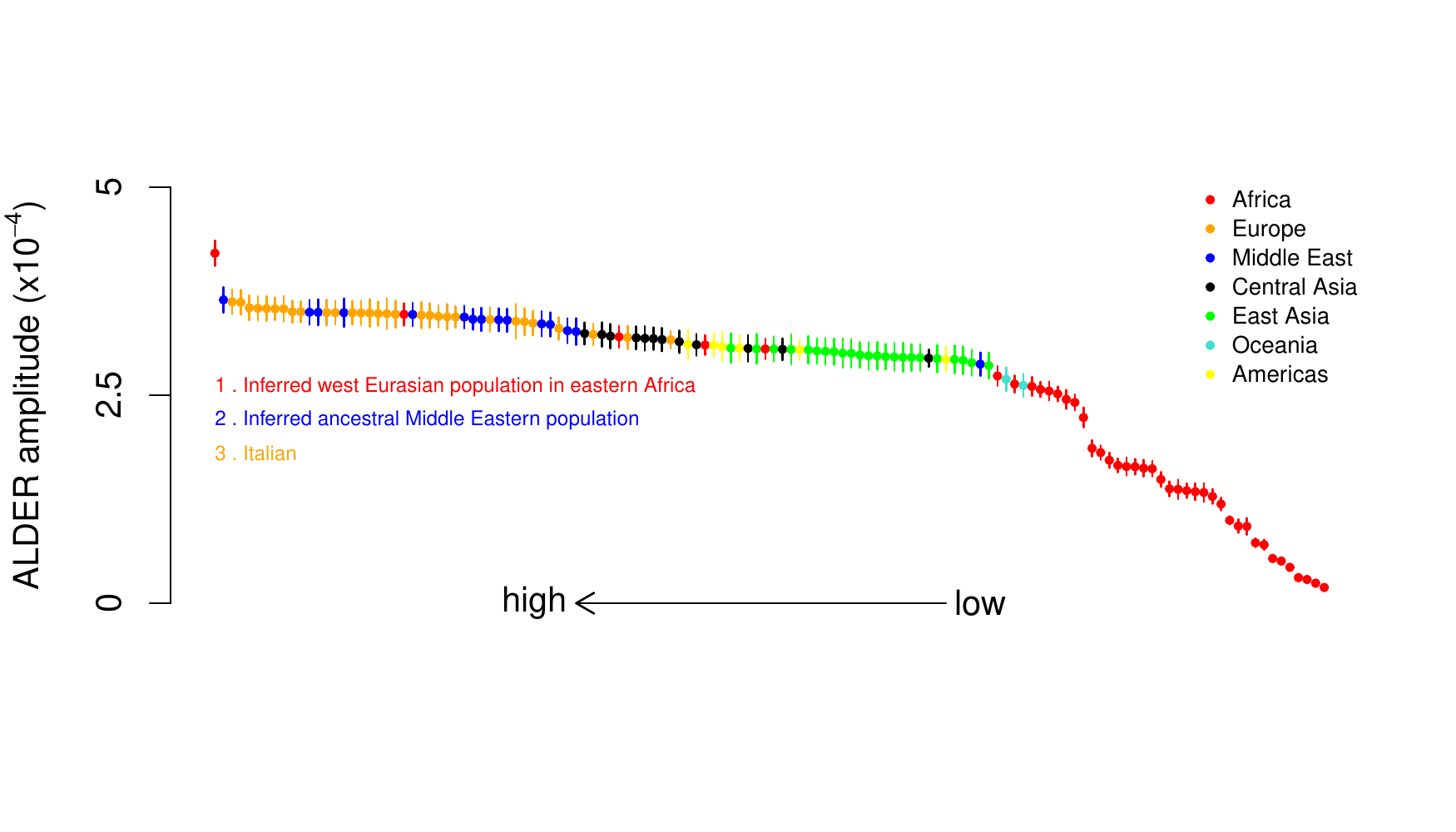}
\caption{\textbf{: Inferring the source of west Eurasian ancestry in southern Africa.} We computed weighted LD curves in the Juhoansi using the Juhoansi as one reference and a range of other populations as the second reference. We then fitted an exponential decay curve to each LD curve, starting from 0.5 cM. Plotted are the fitted amplitudes for each curve. Error bars indicate one standard error. A larger amplitude indicates a closer relationship to one of the true admixing populations. Populations are ordered according to the amplitude, and colored according to their continent of origin. Included along with sampled populations are two inferred populations: the inferred west Eurasian population that entered Ethiopia, and an inferred Middle Eastern population prior to admixture with African populations \citep{Moorjani:2011ys}.}\label{fig_illumina_alder}
\end{center}
\end{figure}

To test if the west Eurasian ancestry in southern and eastern Africa is from the same source, we reconstructed the allele frequencies of the west Eurasian population involved in the admixture in eastern Africa (SI Methods). We then tested whether this hypothetical population is a good proxy for the west Eurasian ancestry in southern Africa. Indeed, this reconstructed population is a better proxy than samples of modern Eurasians (Figure \ref{fig_illumina_alder}). In the Juhoansi (who correspond to the Ju$|$'hoan\_North), we obtain an ALDER amplitude in the one-reference test of $4.2 \times 10^{-4} \pm 1.5 \times 10^{-5}$ when using this imputed population as a reference versus $3.6 \times 10^{-4} \pm 1.5 \times 10^{-5}$ when using Italians as a reference (one-sided P-value for difference of $0.0015$). 

We then applied our method for dating multiple admixture events to the eastern African populations in these data (Supplementary Figure 25-39). Pagani et al. \citep{Pagani:2012uq} previously dated the earliest admixture events in Ethiopia to around 3,000 years ago, but with considerable variation between populations. We find evidence for multiple episodes of population mixture in eastern Africa; most populations have evidence for an early admixture event that we date to around 80-110 generations (2,400-3,300 years) ago (Figure \ref{fig_illumina}). As in southern Africa, the west Eurasian ancestry is present in the early admixture event (Supplementary Table 6). The earliest dates of population mixture that we estimate in eastern Africa are almost uniformly older than those we estimate in southern Africa (Figure \ref{fig_illumina}). One potential concern regarding this conclusion is that the southern and eastern African populations displayed in Figure \ref{fig_illumina} were genotyped on different genotyping arrays; however, this pattern remains when using only populations typed on the same array (Supplementary Figure 40). We conclude that the west Eurasian ancestry in southern Africa was likely brought by a migration of an already-admixed population from eastern Africa.  

\subsection*{Estimating the proportion of eastern African ancestry in southern Africa.}
If west Eurasian ancestry indeed entered southern Africa via eastern Africa, then the relative proportions of west Eurasian ancestry in different southern African populations can be interpreted as reflecting different levels of eastern African gene flow. We thus attempted to split the ancestry of all southern African populations into three components: Khoisan ancestry, putative eastern African ancestry, and ancestry from Bantu-speaking immigrants to southern Africa. To do this, we make the following assumptions: first, that all eastern African ancestry in southern Africa came from a single source with a fixed level of west Eurasian admixture; and second, that all non-Khoisan ancestry in southern Africa is either from this putative eastern African population or from a Bantu-speaking population. Because these assumptions are largely unverifiable, the following should be viewed as more qualitative than quantitative.

We first attempted to estimate the proportion of west Eurasian ancestry in the putative eastern African population that entered southern Africa. Using ALDER, we estimate the lower bound on the proportion of non-Khoisan ancestry in the Ju$|$'hoan\_North as 4\%. If approximately 1\% of this is west Eurasian ancestry (Table \ref{frac_table}) and the Ju$|$'hoan\_North have no Bantu-related ancestry, then this gives an admixture proportion of $\sim$ 25\% west Eurasian ancestry in the putative eastern African source population. Using this value, we then estimated the proportions of Khoisan, putative eastern African, and Bantu-related ancestry of all populations using a linear model (\citep{Patterson:2010fk}, SI Methods). In Table \ref{saf_table}, we show our estimates of these three components (excluding in this case the Nama, who have recent European ancestry that confounds this analysis).

\section*{Discussion}
In this paper, we have examined the history of southern and eastern African populations using patterns of admixture LD. The most striking inference from this analysis is the presence of west Eurasian ancestry in southern Africa that we date to 900-1,800 years ago. Several lines of evidence suggest that the population that brought this ancestry to southern African was an already-admixed population from eastern Africa. 

\paragraph{Back-to-Africa gene flow in eastern Africa.} 

A major open question concerns the initial source of the west Eurasian ancestry in eastern Africa. The estimated mean time of gene flow in eastern Africa is around 3,000 years ago, and the amount of gene flow must have been quite extensive, as the west Eurasian ancestry proportions reach 40-50\% in some Ethiopian populations (Table \ref{frac_table} and \citep{Pagani:2012uq}). Archaeological records from this region are sparse, so Pagani et al. \citep{Pagani:2012uq} speculate that this admixture is related to the Biblical account of the Kingdom of Sheba. However, archaeological evidence is not completely absent. During this time period, architecture in the Ethiopian culture of D'mt has an ``unmistakable South Arabian appearance in many details" \citep{munro1991}, though there is some debate as to whether these patterns can be attributed to large movements of people versus elite-driven cultural practices \citep{Mitchell:2005uq, munro1991}. Additionally, linguistic evidence suggests that this time period was when Ethiosemitic languages were introduced to Africa, presumably from southern Arabia \citep{Kitchen:2009uq}. It is perhaps not a coincidence that the highest levels of west Eurasian ancestry in eastern Africa are found in the Amhara and Tygray, who speak Ethiosemitic languages and live in what was previously the territory of D'mt and the later kingdom of Aksum.

\paragraph{West Eurasian ancestry in southern Africa.}
A second question is: which population or populations introduced west Eurasian ancestry into southern Africa? The best genetic proxy for this ancestry that we have found is the west Eurasian ancestry in eastern Africa (Figure \ref{fig_illumina_alder}), and though we do not identify modern east African populations as the best source population, this is likely due to the lack of genetic drift specific to eastern Africa (SI Methods, Section 1.2.3). The most parsimonious explanation for this observation is that west Eurasian ancestry entered southern Africa indirectly via eastern Africa (though the alternative scenario of direct contact with an unsampled west Eurasian population cannot formally be excluded; however, there is no archaeological, historical, or linguistic evidence of such contact). The relevant eastern African population may no longer exist. However, such a migration has been suggested based on shared Y chromosome haplotypes \citep{cruciani2002back, henn2008chromosomal} and shared alleles/haplotypes associated with lactase-persistence \citep{coelho2009edge, Schlebusch:2012uq} between the two regions. Furthermore, based on a synthesis of archaeological, genetic, climatological and linguistic data G\"uldemann \citep{gldemann2008linguist} hypothesized that the ancestor of the Khoe-Kwadi languages in southern Africa was brought to the region by immigrating pastoralists from eastern Africa. Our observation of elevated west Eurasian ancestry in Khoe-Kwadi groups in general (Table \ref{frac_table}) is consistent with this hypothesis. 

\paragraph{Alternative historical scenarios.} We note that we have interpreted admixture signals in terms of large-scale movements of people. An alternative frame for interpreting these results might instead propose an isolation-by-distance model in which populations primarily remain in a single location but individuals choose mates from within some relatively small radius. In principle, this sort of model could introduce west Eurasian ancestry into southern Africa via a ``diffusion-like" process. Two observations argue against this possibility. First, the gene flow we observe is asymmetric: while some eastern African populations have up to 50\% west Eurasian ancestry, levels of sub-Saharan African ancestry in the Middle East and Europe are considerably lower than this (maximum of 15\% \citep{Moorjani:2011ys}) and do not appear to consist of ancestry related to the Khoisan. Second, the signal of west Eurasian ancestry is present in southern Africa but absent from central Africa, despite the fact that central Africa is geographically closer to the putative source of the ancestry. These geographically-specific and asymmetric dispersal patterns are most parsimoniously explained by migration from west Eurasia into eastern Africa, and then from eastern to southern Africa. 

\paragraph{Conclusions.} Based on these analyses, we can propose a model for the spread of west Eurasian ancestry in southern and eastern Africa as follows: first, a large-scale movement of people from west Eurasia into Ethiopia around 3,000 years ago (perhaps from southern Arabia and associated with the D'mt kingdom and the arrival of Ethiosemitic languages) resulted in the dispersal of west Eurasian ancestry throughout eastern Africa. This was then followed by a migration of an admixed population (perhaps pastoralists related to speakers of  Khoe-Kwadi languages) from eastern Africa to southern Africa, with admixture occurring approximately 1,500 years ago. Advances in genotyping DNA from archaeological samples may allow aspects of this model to be directly tested.

\section*{Acknowledgements}

We thank Richard Durbin for suggesting that we impute the allele frequencies of the ancestral west Eurasian population in eastern Africa, and Carlos Bustamante for helpful discussions on the interpretation of the admixture signals. We thank two anonymous reviewers and commenters on the blogs Gene Expression and Haldane's Sieve for helpful suggestions. The Khoisan samples analysed here were collected in the framework of a multidisciplinary project investigating the prehistory of the Khoisan peoples and languages (http://www2.hu-berlin.de/kba/), and we thank Tom GŸldemann, Christfried Naumann, Chiara Barbieri, Linda Gerlach, Falko Berthold, and Hirosi Nakagawa for assistance with sample collection. JP was supported by NIH postdoctoral fellowship GM103098. This work, as part of the European Science Foundation EUROCORES Programme EuroBABEL, was supported by a grant from the Deutsche Forschungsgemeinschaft (to BP) and by funds from the Max Planck Society (to BP and MS). M.L. and P.L. acknowledge NSF Graduate Research Fellowship support. P.L. was also partially supported by NIH training grant 5T32HG004947-04 and the Simons Foundation. N.P. and D.R. were funded by NIH grant GM100233 and NSF HOMINID grant 1032255

\clearpage

\begin{table}[h]
\caption{\textbf{Estimates of the proportion of west Eurasian ancestry. A In southern African populations. B In eastern African populations.} We estimated the percentage of west Eurasian ancestry in each southern and eastern African population (see SI Appendix).  For each region, populations are sorted according to the estimated proportion of west Eurasian ancestry. Standard errors on all estimates ranged from 0.3\% to 1.1\%, with an average of 0.7\%. We additionally estimated the Eurasian ancestry proportion in the Mandenka from western Africa using this method as 2.0 \%.} 
\label{frac_table}

\subtable[Southern Africa]{
\begin{tabular}{ l | c | c|  c|}
Population & Language &  West Eurasian \\ 
& classification  & ancestry (\%) \\ \hline
Nama & Khoe-Kwadi  & 14.0\\ \hline
Shua & Khoe-Kwadi  &5.4 \\  \hline
Hai$||$om& Khoe-Kwadi  & 5.2 \\ \hline
 Khwe& Khoe-Kwadi & 4.0 \\ \hline
 Tshwa & Khoe-Kwadi & 3.0 \\ \hline
Naro & Khoe-Kwadi  & 2.2 \\ \hline
G$|$ui & Khoe-Kwadi & 2.0 \\ \hline
Taa\_North & Tuu &  1.9 \\ \hline
G$||$ana & Khoe-Kwadi &1.6 \\ \hline
!Xuun  & Kx'a & 1.2 \\ \hline
\textdoublebarpipe Hoan & Kx'a & 1.5 \\ \hline
Damara & Khoe-Kwadi & 1.3 \\ \hline
Kgalagadi & Bantu & 1.1 \\ \hline
Ju$|$'hoan\_North& Kx'a & 1.0\\ \hline
Taa\_East & Tuu &  0.4 \\ \hline
Mbukushu & Bantu&0.5 \\ \hline
 Taa\_West &Tuu  & 0.4 \\ \hline
Himba& Bantu  &0.1 \\ \hline
Ju$|$'hoan\_South & Kx'a  & 0 \\ \hline
Tswana & Bantu & 0 \\ \hline
Wambo & Bantu  &0 \\ \hline
\end{tabular}
}

\subtable[Eastern Africa]{
\begin{tabular}{ l | c | c|  c|}
Population & Language &  West Eurasian \\ 
& classification &  ancestry (\%) \\ \hline
Tygray & Semitic &  50.4 \\ \hline
Amhara & Semitic &  49.2 \\ \hline
Afar & Cushitic &  46.0 \\ \hline
Oromo & Cushitic &  41.6 \\ \hline
Somali & Cushitic &  38.4 \\ \hline
Ethiopian Somali & Cushitic & 37.9 \\ \hline
Wolayta & Omotic &  34.1 \\ \hline
Maasai & Nilotic &  18.9 \\ \hline
Ari Cultivator & Omotic &  18.2 \\ \hline
Sandawe & isolate &  15.8 \\ \hline
Ari Blacksmith & Omotic &  15.7 \\ \hline
Hadza & isolate &  6.4 \\ \hline

Luhya & Bantu &  2.4 \\ \hline
Gumuz & isolate &   1.7 \\ \hline
Anuak & Nilotic &  0 \\ \hline
\end{tabular}
}

\end{table}

\clearpage

\begin{table}[!tbp]
\caption{: \textbf{Estimates of the proportion of Khoisan, putative eastern African, and putative Bantu-related ancestry in southern African populations, ordered by the amount of putative eastern African ancestry.} The Nama were excluded from this analysis because of their recent European ancestry. Additionally shown is the proportion of west Eurasian ancestry in each population as estimated by the linear model (these proportions are slightly different from those in Table \ref{frac_table}). }
\label{saf_table}
\begin{center}
\begin{tabular}{ l | c | c| c|}
Population & Khoisan & Putative eastern African (west Eurasian) & Putative Bantu-related \\ 
& ancestry (\%) & ancestry (\%) & ancestry (\%) \\ \hline
Hai$||$om& 54 & 25 (6.3) & 21 \\ \hline
Shua & 37 &21 (5.2)& 43 \\  \hline
Khwe& 36 & 18 (4.6) & 45 \\ \hline
G$|$ui & 80 & 13 (3.2)& 6 \\ \hline
Tshwa & 48 & 10 (2.4) & 43 \\ \hline
!Xuun  & 73 & 9 (2.2) & 18 \\ \hline
Naro & 87 & 9 (2.2)& 5 \\ \hline
Taa\_North & 84 & 9 (2.4) & 7 \\ \hline
G$||$ana & 53 & 6 (1.5)&41 \\ \hline
\textdoublebarpipe Hoan & 70 & 6 (1.4)& 24 \\ \hline
Ju$|$'hoan\_South & 93 &6 (1.5)& 1 \\ \hline
Damara & 9 & 4 (1.0) & 88 \\ \hline
Ju$|$'hoan\_North& 96 & 4 (1.0)  & 0\\ \hline
Mbukushu & 9& 2 (0.5)&89 \\ \hline
Taa\_East & 74 & 1 (0.2)& 25 \\ \hline
Taa\_West &83 &1 (0.3)& 16 \\ \hline
Himba& 8 & 0 (0)& 92 \\ \hline
Tswana & 22 &0 (0)& 78 \\ \hline
Kgalagadi & 33 & 0 (0)&67 \\ \hline
Wambo & 5 &0 (0)& 95 \\ \hline
\end{tabular}
\end{center}
\label{saf_table}
\end{table}
\clearpage

\bibliography{../../bib}



\section*{Supplementary material}

# Supplementary information for: Ancient west Eurasian ancestry in southern and eastern Africa

Joseph K. Pickrell[1,†], Nick Patterson[2], Po-Ru Loh[3], Mark Lipson[3], Bonnie Berger[2,3], Mark Stoneking[4], Brigitte Pakendorf[5,†], David Reich[1,2,†]

[1] Department of Genetics, Harvard Medical School
[2] Broad Institute
[3] Department of Computer Science and Mathematics and AI Laboratory, MIT
[4] Department of Evolutionary Genetics, MPI for Evolutionary Anthropology
[5] Laboratoire Dynamique du Langage, UMR5596, CNRS and Université Lyon Lumière 2
[†] To whom correspondence should be addressed: joseph_pickrell@hms.harvard.edu, brigitte.pakendorf@cnrs.fr, reich@genetics.med.harvard.edu

October 21, 2013

# Contents

# 1 Materials and Methods

## 1.1 Details of analyzed samples

### 1.1.1 Affymetrix data

The main southern African dataset in this paper is from Pickrell et al. [2012] (the dataset before filtering for outlier individuals). We genotyped an additional 32 samples from these populations on the Affymetrix Human Origins Array (Supplementary Table 1). As with the original dataset [Pickrell et al., 2012], these additional samples were obtained as part of a multidisciplinary project investigating the prehistory of the Khoisan peoples and languages (http://www2.hu-berlin.de/kba/) with prior informed written consent from all donors, and with ethical clearance from the Review Board of the University of Leipzig and with the permission of the Ministry of Youth, Sport and Culture of Botswana and the Ministry of Health and Social Services of Namibia. We then examined all populations in this merged dataset for outliers, and removed 17 individuals (Supplementary Figure 1).

We merged these data with those from Patterson et al. [2012] and Meyer et al. [2012], and all other samples genotyped on the same array. In analyses of multiple mixture events, we used a set of 51 populations (Supplementary Table 2). For most analyses of the Khoisan, we excluded the Damara because they appear genetically similar to Niger-Congo speakers [Pickrell et al., 2012]. These data consist of 565,259 SNPs; for most analyses, we use all of these SNPs. However, in some places (where noted) we used only subsets of these SNPs from known ascertainment panels. For analysis of multiple mixture dates we used the set of populations listed in Supplementary Table 2.

### 1.1.2 Illumina data

We merged data from several published sources [Altshuler et al., 2010; Behar et al., 2010; Henn et al., 2011; Li et al., 2008; Pagani et al., 2012; Schlebusch et al., 2012]. The merged dataset consisted of 2,935 individuals genotyped at 256,540 SNPs. For analyses of multiple mixture events, we used a set of 55 populations (Supplementary Table 7).

## 1.2 Estimating multiple dates of population mixture from weighted LD.

### 1.2.1 Theory

Here, we consider the properties of admixture LD in the presence of multiple admixture events in the history of a population. Consider two bi-allelic SNPs, $x$ and $y$, in a haploid population $T$, and let the covariance between the genotypes (coded as 0 and 1 according to an arbitrary reference) be $z_T(x, y)$. This follows the notation in Loh et al. [2013]; note that $z_T(x, y)$ is simply the standard measure of LD often called D. The demographic history of population $T$ influences $z_T(x, y)$ in a known fashion. First, if $T$ derived from a single admixture event between two populations $A$ and $B$ with mixture proportions of $\alpha$ and $1 - \alpha$, respectively, then $t_1$ generations after admixture in a population of infinite size [Chakraborty and Weiss, 1988]:

$$z_T(x,y) = [\alpha z_A(x,y) + (1-\alpha)z_B(x,y) + \alpha(1-\alpha)\delta_{AB}(x)\delta_{AB}(y)]e^{-t_1 d}, \qquad (1)$$

where $d$ is the genetic distance between $x$ and $y$ and $\delta_{AB}(x)$ is the difference in allele frequencies at SNP $x$ between populations $A$ and $B$. In Loh et al. [2013], it is assumed that $z_A(x, y) = z_B(x, y) = 0$. However, instead consider the case where population $A$ itself is descended from admixture between populations $C$ and

$D$ at time $t_2$, with admixture fractions $\beta$ and $1 - \beta$, respectively (Supplementary Figure 4). If $z_C(x, y) = z_D(x, y) = 0$ (i.e. neither $C$ nor $D$ is admixed), then:

$$z_T(x, y) = [\alpha\beta(1-\beta)\delta_{CD}(x)\delta_{CD}(y)e^{-(t_2-t_1)d} + \alpha(1-\alpha)\delta_{AB}(x)\delta_{AB}(y)]e^{-t_1 d} \quad (2)$$

$$= \alpha\beta(1-\beta)\delta_{CD}(x)\delta_{CD}(y)e^{-t_2 d} + \alpha(1-\alpha)\delta_{AB}(x)\delta_{AB}(y)e^{-t_1 d}. \quad (3)$$

If $C$ or $D$ is itself admixed, this simply adds another exponential term to $z_T(x, y)$. Generalizing to $n$ population mixtures, we can see that

$$z_T(x, y) = \sum_{i=1}^{n} W_i \delta_i(x) \delta_i(y) e^{-t_i d}, \quad (4)$$

where $t_1, t_2, ..., t_n$ are the times of the various mixture events, $W_i$ is a function of the mixture proportions of event $i$ (and the contribution of this mixture event to $T$), and $\delta_i(x)$ is the difference in allele frequencies at locus $x$ between the two populations involved in mixture event $i$.

Now, following Loh et al. [2013], consider two reference populations, $A'$ and $B'$. We can now define a weighted measure of LD:

$$a_{A'B'}(x, y) = z_T(x, y)\delta_{A'B'}(x)\delta_{A'B'}(y) \quad (5)$$

$$= \sum_{i=1}^{n} W_i \delta_i(x) \delta_{A'B'}(x) \delta_i(y) \delta_{A'B'}(y) e^{-t_i d}. \quad (6)$$

If we then take the expected value of $a_{A'B'}$ at some genetic distance $d$:

$$E[a_{A'B'}(d)] = \sum_{i=1}^{n} W_i E[\delta_i \delta_{A'B'}]^2 e^{-t_i d}. \quad (7)$$

In the diploid case, all the entries here are simply multiplied by a factor of two [Loh et al., 2013].

### 1.2.2 Model fitting

Consider a set of $m$ reference populations, $X_1, X_2, ..., X_m$, from which we have sampled $N_1, N_2, ..., N_m$ individuals, respectively, and genotyped $L$ SNPs. Let there be a single target population $T$, and we have sampled $N_T$ samples from this population. We can calculate the weighted LD statistic in population $T$ using each pair of reference populations $i$ and $j$:

$$\hat{a}_{ij}(d) = \frac{\sum_{\{x,y\} \in S(d)} \hat{z}_T(x, y) \hat{\delta}_{ij}(x) \hat{\delta}_{ij}(y)}{|S(d)|}, \quad (8)$$

where $\hat{\delta}_{ij}(x) = \hat{f}_i(x) - \hat{f}_j(x)$, $\hat{f}_i(x)$ is the trivial estimator of the allele frequency at locus $x$ in population $i$, $S(d)$ is the set of all pairs of SNPs separated by genetic distance $d$, and

$$\hat{z}_T(x, y) = \frac{1}{N_T - 1} \sum_{k=1}^{N_T} (g_{kx} - \bar{g}_x)(g_{ky} - \bar{g}_y), \quad (9)$$

where $g_{kx}$ is the genotype of individual $k$ in population $T$ at locus $x$ (coded as 0, 1, or 2 copies of an

3

arbitrarily defined reference allele), and $\bar{g}_x = \frac{1}{N_T} \sum_k g_{kx}$.

For a given target population, then, we can calculate $\binom{m}{2}$ curves of weighted LD (in practice, we can do this extremely quickly using the algorithm in Loh et al. [2013]). The theory above tells us that each curve is a mixture of exponential curves. We thus model each curve as:

$$\hat{a}_{ij}(d) = K_{ij} + \sum_{k=1}^{n} C_{ijk} e^{-t_k d} + \epsilon_{ij}(d), \tag{10}$$

where $K_{ij}$ is an affine term estimated for each pair of populations, $C_{ijk}$ is the amplitude of the $k$th exponential term for populations $i$ and $j$, $\epsilon_{ij}(d)$ are error terms distributed as $N(0, \sigma_{ij}^2)$ (note that these error terms are not independently distributed, so we will use a jackknife to judge fit), and $t_k$ is the time of the $k$th admixture event. *The key fact is that different pairs of reference populations often have different relative values of $C_{ij}$ but always have fixed values of t.* This in principle gives us some leverage in the tricky problem of fitting mixtures of exponentials.

We now want to estimate all the parameters in Equation 10. These include the number of waves of mixture, the amplitudes, and the admixture times. We treat this as a least squares problem; that is, we want to minimize $\sum_{ij} (\hat{a}_{ij}(d) - E[a_{ij}(d)])^2$. We start by assuming a single wave of admixture. For a fixed time, the amplitudes can be solved by non-negative least squares. We then numerically optimize the admixture times (also enforcing non-negativity) using the Nelder-Mead algorithm implemented in the GNU scientific library [Galassi et al., 2002]. Once the model is fit, we calculate jackknife standard errors of all the parameters, by dropping each chromosome in turn and re-optimizing. In all cases, we started fitting curves only from 0.5 cM. If the curve is "significant" we add another exponential term. In total, the algorithm is:

1. Add a new exponential term to the model.
2. Fit the model by alternately optimizing all exponential decay terms and amplitudes.
3. Calculate standard errors on all terms using a jackknife.
4. If all decay terms have a p-value less than 0.01, go back to step 1, otherwise finish.

In this model, each pair of populations is treated as independent. We thus additionally experimented with performing a bootstrap where we randomly sample pairs of populations rather than re-sampling chromosomes. The results from this analysis were qualitatively similar to those presented, so we use the standard errors from the above jackknife procedure.

To infer the sources of admixture for each admixture time, we examined the $C_{ijk}$ parameters (recall that these are the amplitudes of the LD curve computed using populations $i$ and $j$ on the admixture time $k$). For each Khoisan population, we identified the maximum $C_{ijk}$ where $i$ or $j$ was a Niger-Congo-speaking group and the other was a Khoisan group. Call this $C_{NC}^{max}$. We then identified the maximum $C_{ijk}$ where either $i$ or $j$ was a west Eurasian group and the other was a Khoisan group. Call this $C_{WE}^{max}$. We also have the standard errors on these estimates from the jackknife. We then computed a Z-score to test whether these were significantly different:

$$Z = \frac{C_{NC}^{max} - C_{WE}^{max}}{\sqrt{se(C_{NC}^{max})^2 + se(C_{WE}^{max})^2}}. \tag{11}$$

If the p-value from this test was greater than 0.05, in Figure 4 in the main we show this as a low-confidence ancestry call.

To compare the amplitudes in the eastern African populations, we performed the same type of test. The exact same test is not possible because we have no set of populations in eastern Africa that are analogous to

the Khoisan in southern Africa in representing presumably autochthonous groups. Instead, we compared the set of $C_{ijk}$ where one reference population was Eurasian to the set of $C_{ijk}$ where neither reference population was Eurasian. We separately compared the set of $C_{ijk}$ where one population was a Niger-Congo-speaking agriculturalist group to the set of $C_{ijk}$ where neither population was a Niger-Congo-speaking agriculturalist group. We then report the relevant Z-score (that is, if the overall maximum $C_{ijk}$ included a Eurasian population, we reported the first Z-score, while if the overall maximum $C_{ijk}$ included a Niger-Congo-speaking agriculturalist group, we reported the second Z-score. In the cases where the overall maximum $C_{ijk}$ included both a Eurasian population and a Niger-Congo-speaking population, we reported the minimum of the two Z-scores).

### 1.2.3 Conditions under which the mixture weights do not identify the true admixing populations.

When fitting the above model, for a given pair of populations $i$ and $j$ on admixture event $k$, we estimate a parameter $C_{ijk}$. The relative values of these parameters for different populations reflect the differential relationships of $i$ and $j$ to the true admixing populations. We asked in what situation the maximum values of these parameters do not identify the reference populations most closely related to the true admixing populations.

Consider the two-admixture situation presented in Supplementary Figure 4, and let the earliest admixture event (between populations $C$ and $D$) be too old to detect by LD, and so the only curve in the data has a decay rate of $t_2$ generations. Now consider the LD curve computed using populations $A$ and $B$ as references (these are the true admixing populations) and that computed using $B$ and $C$ as references. The two curves have the form:

$$E[a_{AB}(d)] = \alpha(1-\alpha)E[\delta_{AB}\delta_{AB}]^2 e^{-t_2 d} \tag{12}$$
$$E[a_{CB}(d)] = \alpha(1-\alpha)E[\delta_{AB}\delta_{CB}]^2 e^{-t_2 d}, \tag{13}$$
$$\tag{14}$$

so the curve computed using the true mixing populations will have the highest amplitude when $E[\delta_{AB}\delta_{AB}]^2 > E[\delta_{AB}\delta_{CB}]^2$. Writing these out explicitly (using the branch lengths in Supplementary Figure 4),

$$E[\delta_{AB}\delta_{AB}] = x_1 + x_2 + \beta^2 x_3 + (1-\beta)^2 x_4 + x_5 \tag{15}$$
$$E[\delta_{CB}\delta_{AB}] = x_1 + x_2 + \beta x_3. \tag{16}$$

Thus, weighted LD curves computed using the true mixing populations have the highest amplitude when $\beta^2 x_3 + (1-\beta)^2 x_4 + x_5 > \beta x_3$. Note that $x_3$ is weighted by $\beta^2$ on the left hand side and $\beta$ on the right hand side. In our applications, European populations correspond to population $C$, $\beta$ corresponds to the amount of west Eurasian ancestry, and the branch $x_3$ corresponds to the out-of-Africa bottleneck, which induced a large amount of genetic drift. We thus expect the amplitude to be dominated by the $x_3$ term, and to identify a Eurasian population as the best reference population if the true source population has even a low level of Eurasian admixture.

### 1.2.4 Simulations

To test the performance of our approach to estimate multiple mixture dates, we used coalescent simulations implemented in the software macs [Chen et al., 2009]. The basic simulation setup is shown in Supplementary Figure 5. For each simulation, we simulated 30 individuals from each of the nine populations, and each individual consisted of 10 independent chromosomes of 200Mb. We thus simulated many aspects of real data, including hundreds of thousands of SNPs in linkage disequilibrium. We simulated three scenarios: $t_1 = 20$ and $t_2 = 100$, $t_1 = 20$ and $t_2 = 60$, and $t_1 = 40$ and $t_2 = 100$. We additionally performed simulations of a single episode of population mixture. The exact macs command used was (for, e.g., $t_1 = 40$ and $t_2 = 100$):

```
macs 540 200000000 -t 0.00004 -r 0.0004 -I 9 60 60 60 60 60 60 60 60 60 0 -em 0.0010 1 4 8000 -em 0.001025 1 4 0 -em 0.0025 1 7 8000 -em 0.002525 1 7 0 -ej 0.0125 7 8 -ej 0.0125 1 2 -ej 0.0125 4 5 -en 0.0249 8 0.02 -ej 0.025 8 9 -en 0.0249 2 0.02 -ej 0.025 2 3 -en 0.0249 5 0.02 -ej 0.025 5 6 -en 0.0748 9 0.01 -ej 0.075 6 9 -en 0.1498 9 0.01 -ej 0.15 3 9
```

Each simulation consists of 10 replicates of the above command. We then ran our method to identify multiple mixture dates on the admixed population using all eight other simulated populations as references. Results are shown in Supplementary Figure 6. In the case with a large difference in the mixture times ($t_1 = 20$ and $t_2 = 100$), the method behaves well in all simulations. In harder cases (when either the mixture times are close together or both are old), the method occasionally misses the second mixture event or results in extremely large confidence intervals. Additionally, in two simulations where the method identifies the correct number of admixture events, the confidence intervals on the dates do not overlap the true values. Overall, however, in 21/23 such simulations (where the method identifies the correct number of admixture events), the 95% confidence intervals include the true values.

To gain some intuition into the performance of the model, we examined the LD curves underlying the fitted models. In Supplementary Figure 8, we show an example fitted model under the simulation where $t_1 = 20$ and $t_2 = 100$, and in Supplementary Figure 9 we show the same plot for a simulation where $t_1 = 40$ and $t_2 = 100$. The reason for the change in performance is clear: when both admixture events are older, the fit of the single exponential curve is already reasonably good and adding a second exponential in this case only marginally (though significantly in this case) improves the fit. However, when one of the admixture events is more recent, the fit of the single exponential curve becomes visibly poor.

We additionally simulated a situation with a low level of gene flow from population 4 into population 1 over the course of 100 generations (rather than in a single pulse, as assumed in the model). The exact macs command used for this simulation was:

```
macs 540 200000000 -t 0.00004 -r 0.0004 -I 9 60 60 60 60 60 60 60 60 60 0 -em 0.0005 1 4 80 -em 0.003 1 4 0 -ej 0.0125 7 8 -ej 0.0125 1 2 -ej 0.0125 4 5 -en 0.0249 8 0.02 -ej 0.025 8 9 -en 0.0249 2 0.02 -ej 0.025 2 3 -en 0.0249 5 0.02 -ej 0.025 5 6 -en 0.0748 9 0.01 -ej 0.075 6 9 -en 0.1498 9 0.01 -ej 0.15 3 9
```

Again, each simulation consisted of 10 replicates of the above command, and applied our model. In this situation, the model sometimes infers that this represents two pulses of admixture from the same source (Supplementary Figure 7). In applications to real data, we sometimes infer two pulses of admixture from the same source (e.g. in the Nama and Taa_West in southern Africa, and the Oromo, Maasai, Amhara, and Gumuz in eastern Africa); in these cases, a potential alternative interpretation is low level gene flow over a long period of time rather then two discrete pulses of gene flow.

An additional source of error for methods like this are demographic complexities like population bottlenecks. Loh et al. [2013] show that a shared bottleneck in the history of one of the reference populations and a

test population can generate a curve of weighted LD that mimics that of admixture to some degree. However, the amplitudes of bottleneck-induced weighted LD curves do not depend on the reference populations used [Loh et al., 2013], so we do not consider this further in our analyses of southern Africa.

## 1.3 Analysis of combined populations

In most southern African populations where our method detects only a single admixture event, the fitted model visually appears inadequate to fully explain the data (e.g. Supplementary Figures 10, 12, 17, 20). Indeed, there is marginal statistical evidence for two admixture events in many of these populations (Supplementary Table 3). As described in the main text, we performed analyses of combined populations. Note that combining populations induces LD across the whole genome, but does not induce a decay curve; if populations share an admixture event, combining the populations should result in increased power to detect it.

We first combined a set of populations (the Tshwa, Shua, Hai||om, ǂHoan, Naro, and Taa_North) that appear to have weak evidence for a second, more recent admixture event (Supplementary Table 3), and ran our method on this combined sample. In this combined sample we infer two dates of admixture: one $40 \pm 2$ generations ago and one $4 \pm 1$ generations ago (Z-score of 3.2, $P = 7 \times 10^{-4}$), with the more recent admixture involving Bantu-speaking populations, though this ancestry assignment is made with low confidence.

We then combined two populations (the Ju|'hoan_North and G|ui) that have weak evidence for a second, more ancient admixture event (Supplementary Table 3). In this combined sample (Supplementary Figure 23), we also infer two dates of admixture, but with different dates from all other samples: one $30 \pm 4$ generations ago, and one $109 \pm 41$ generations ago (Z-score of 2.6, $P = 0.005$). We interpret this as evidence that the population that introduced west Eurasian ancestry to southern Africa was itself admixed, and that this more ancient admixture happened around 110 generations ago (though the confidence intervals here are clearly large). However, we cannot exclude the possibility that the Ju|'hoan_North and G|ui alone experienced gene flow from this admixed population, while the west Eurasian ancestry detected in the other southern African populations stems from a different population that did not carry this signal of ancient admixture.

## 1.4 $f_4$ ancestry estimation.

To estimate the fraction of west Eurasian ancestry in each African population, we used the fact that this ancestry appears to be more closely related to southern Europe and the Middle East than to northern Europe. We thus computed the $f_4$ ratio $f_4$(Han, Orcadian; X, Druze) / $f_4$(Han, Orcadian; Yoruba, Druze), where X is any African population. (Supplementary Figure 40). Since the Druze have a small level of west African ancestry, this ratio is not exactly the desired fraction. Instead, if we let $\lambda$ be the fraction of Druze-like ancestry in population X and $F$ be the fraction of Yoruba-like ancestry in the Druze:

$$\frac{f_4(H,O;X,D)}{f_4(H,O;Y,D)} = \frac{1-\lambda-F}{1-F}. \tag{17}$$

We approximate $\lambda$ by assuming $F = 0.05$ for the Druze [Moorjani et al., 2011]. In some cases, our estimates of west Eurasian ancestry are slightly below zero (though not statistically significantly so); for these populations we report the ancestry proportion as 0%.

7

## 1.5 Estimating the allele frequencies of the west Eurasian population that admixed into eastern Africa.

We sought to impute the allele frequencies of the ancestral west Eurasian population that entered eastern Africa. To do this, we model the allele frequencies in a set of $N$ eastern African populations as a weighted combination of allele frequencies in the Sudanese (we choose the Sudanese because they are often the best proxy for the African ancestry in Ethiopian populations; Supplementary Table 4) and an unknown west Eurasian population. Let $\hat{f}_S$ be the estimated allele frequency at a given SNP in the Sudanese, $f_X$ be the (unknown) allele frequency at the SNP in the ancestral west Eurasian population, and $f_j$ be the population allele frequency (as opposed to the sample allele frequency) of the SNP in eastern African population $j$. We model $f_j$ as:

$$f_j = f_X \hat{w}_X + \hat{f}_S \hat{w}_S, \tag{18}$$

where $\hat{w}_X$ is the estimated proportion of west Eurasian ancestry in population $j$ from Table 1 in the main text and $\hat{w}_S = 1 - \hat{w}_X$. For a given $f_X$, the sum of squared errors is:

$$SS(f_X) = \sum_{j=1}^{N} (\hat{f}_j - f_j)^2, \tag{19}$$

where $\hat{f}_j$ is the estimated allele frequency at the SNP in population $j$. We then search over values of $f_X$ to minimize Equation 19.

In the set of east African populations, we included the six with the most west Eurasian ancestry: the Tygray, Amhara, Afar, Oromo, Somali, and Ethiopian Somali. We then minimized Equation 19 for each SNP in the Illumina data using the optimize() function in R [R Development Core Team, 2011]. To compare to an ancestral Middle Eastern population, we performed the same analysis on the Bedouin, Druze, and Palestinian populations, using the estimated African ancestry proportions from Moorjani et al. [2011].

To run ALDER using these imputed allele frequencies, we calculated the ALDER weighted LD statistic in the Juhoansi (since the eastern African populations were typed on an Illumina array, these are the samples from Schlebusch et al. [2012]) using weights calculated from the Juhoansi as one reference and the imputed allele frequencies as the other reference. To account for sampling error, we simulated 40 individuals from the inferred ancestral west Eurasian population using the estimated allele frequencies.

## 1.6 Partitioning non-Khoisan ancestry into putative eastern African and putative Bantu-related ancestry.

To partition the ancestry of all southern African groups into Khoisan, putative eastern African, and putative Bantu-related ancestry, we model the allele frequency at a given SNP in a Khoisan population $f_X$ as a linear combination of the allele frequencies in the Ju|'hoan_North ($f_J$), Yoruba ($f_Y$), Dinka ($f_D$), and Italian ($f_I$):

$$f_X = w_1 f_J + w_2 f_Y + w_3 f_D + w_4 f_I, \tag{20}$$

where $\sum_{i=1}^{4} w_i = 1$. Using all SNPs, we estimated these weights using the approach of Patterson et al. [2010], using the Han as an outgroup population. If these four populations were the true unadmixed reference populations, these weights would correspond to the mixture fractions in population $X$. Since the Ju|'hoan_North are admixed and the Dinka may not be the best reference for an ancestral east African population, we took the following approach to convert these weights to the admixture fractions we are interested in: First, define

the proportion of west Eurasian ancestry as $w_E = 0.01w_1 + w_4$ and the proportion of Khoisan ancestry as $w_K = 0.96w_1$. We then computed the proportion of putative east African ancestry as $w_{EA} = 4w_E$ and the proportion of putative Bantu-related ancestry as $w_B = 1 - w_K - w_{EA}$.

# 2 Supplementary Tables

| Population | # individuals |
|---|---:|
| Taa_East | 3 |
| Taa_North | 5 |
| Taa_West | 9 |
| G|ui | 6 |
| G||ana | 4 |
| !Xuun | 3 |
| Nama | 2 |

Table 1: **Additional southern African samples typed in this study.**

| Population | # individuals |
|---|---|
| Adygei | 15 |
| Balochi | 21 |
| BantuKenya | 10 |
| BantuSouthAfrica | 6 |
| Basque | 20 |
| Bedouin | 38 |
| BiakaPygmy | 20 |
| Melanesian | 9 |
| Dai | 10 |
| Damara | 13 |
| Dinka | 7 |
| Druze | 32 |
| Taa_East | 8 |
| French | 26 |
| G‖ana | 7 |
| G|ui | 7 |
| Hadza | 24 |
| Hai‖om | 9 |
| Himba | 5 |
| Han | 32 |
| ǂHoan | 7 |
| Italian | 11 |
| Japanese | 28 |
| Ju|'hoan_North | 21 |
| Ju|'hoan_South | 6 |
| Kalash | 18 |
| Khwe | 10 |
| Tshwa | 9 |
| Mandenka | 20 |
| Mbukushu | 5 |
| MbutiPygmy | 11 |
| Mozabite | 25 |
| Nama | 18 |
| Naro | 9 |
| Taa_North | 9 |
| Orcadian | 13 |
| Wambo | 5 |
| Palestinian | 34 |
| Russian | 22 |
| Sardinian | 26 |
| Shua | 9 |
| Tswana | 5 |
| Tuscan | 7 |
| Taa_West | 16 |
| !Xuun | 13 |
| Yoruba | 21 |

Table 2: **Populations typed on the Affymetrix Human Origins array used in analyses of multiple mixture events.**

| Population | Mixture date in generations | Best west African amplitude (populations) | Best Eurasian amplitude (populations) | Z-score (P) |
|---|---|---|---|---|
| ǂHoan | 14 | $3.2 \times 10^{-4} \pm 2.3 \times 10^{-5}$ (Taa_North; Yoruba) | $3.5 \times 10^{-4} \pm 2.8 \times 10^{-5}$ (Taa_North; Druze) | 0.87 (0.38) |
| Tshwa | 23 | $3.6 \times 10^{-4} \pm 2.0 \times 10^{-5}$ (Ju|'hoan_South; Yoruba) | $4.3 \times 10^{-4} \pm 2.6 \times 10^{-5}$ (Italian;Ju|'hoan_South) | 2.02 (0.04) |
| Taa_West | 62 | $1.4 \times 10^{-4} \pm 2.2 \times 10^{-5}$ (Ju|'hoan_South; BantuKenya) | $2.9 \times 10^{-4} \pm 2.5 \times 10^{-5}$ (Sardinian;Ju|'hoan_South) | 4.6 ($4 \times 10^{-6}$) |
| Taa_West | 8 | $1.3 \times 10^{-4} \pm 2.6 \times 10^{-5}$ (Taa_North;Yoruba) | $1.4 \times 10^{-4} \pm 2.5 \times 10^{-5}$ (Taa_North;Kalash) | 0.04 (0.97) |
| Taa_North | 30 | $2.5 \times 10^{-4} \pm 1.4 \times 10^{-5}$ (Taa_West;Yoruba) | $3.9 \times 10^{-4} \pm 2.5 \times 10^{-5}$ (Taa_West;Druze) | 5.2 ($2 \times 10^{-7}$) |
| Taa_East | 28 | $3.0 \times 10^{-4} \pm 1.9 \times 10^{-5}$ (Taa_North;Yoruba) | $3.8 \times 10^{-4} \pm 1.4 \times 10^{-5}$ (Taa_West;Italian) | 3.5 ($5 \times 10^{-4}$) |
| Taa_East | 28 | $8.2 \times 10^{-5} \pm 1.4 \times 10^{-5}$ (G|ui;Yoruba) | $7.1 \times 10^{-5} \pm 1.2 \times 10^{-5}$ (Bedouin;G|ui) | 0.65 (0.52) |
| Shua | 32 | $3.8 \times 10^{-4} \pm 1.6 \times 10^{-5}$ (Ju|'hoan_South;Yoruba) | $4.5 \times 10^{-4} \pm 2.2 \times 10^{-5}$ (Ju|'hoan_South;Druze) | 2.7 (0.007) |
| Naro | 37 | $2.4 \times 10^{-4} \pm 1.4 \times 10^{-5}$ (Ju|'hoan_South;BantuKenya) | $4.9 \times 10^{-4} \pm 2.8 \times 10^{-5}$ (Ju|'hoan_South;Sardinian) | 7.8 ($6 \times 10^{-15}$) |
| Nama | 5 | $1.2 \times 10^{-4} \pm 1.3 \times 10^{-5}$ (Taa_West;Damara) | $4.1 \times 10^{-4} \pm 4.1 \times 10^{-5}$ (Taa_West;Orcadian) | 6.6 ($4 \times 10^{-11}$) |
| Nama | 55 | $2.6 \times 10^{-4} \pm 1.7 \times 10^{-5}$ (Taa_West;BantuKenya) | $6.5 \times 10^{-4} \pm 5.5 \times 10^{-5}$ (Taa_West;Sardinian) | 6.9 ($5 \times 10^{-12}$) |
| Khwe | 60 | $2.0 \times 10^{-4} \pm 1.1 \times 10^{-4}$ (Ju|'hoan_South;Yoruba) | $4.5 \times 10^{-4} \pm 7.1 \times 10^{-5}$ (Ju|'hoan_North;Tuscan) | 2.0 (0.05) |
| Khwe | 17 | $1.7 \times 10^{-4} \pm 8.8 \times 10^{-5}$ (Ju|'hoan_North;Yoruba) | $1.4 \times 10^{-4} \pm 7.9 \times 10^{-5}$ (Ju|'hoan_North;Dai) | 0.3 (0.76) |
| Ju|'hoan_South | 39 | $2.1 \times 10^{-4} \pm 3.1 \times 10^{-5}$ (Ju|'hoan_North;Mandenka) | $3.3 \times 10^{-4} \pm 9.4 \times 10^{-5}$ (Ju|'hoan_North;Dai) | 1.3 (0.19) |
| Ju|'hoan_South | 64 | $5.6 \times 10^{-5} \pm 1.2 \times 10^{-5}$ (G||ana,Mbukushu) | $1.2 \times 10^{-4} \pm 4.1 \times 10^{-5}$ (G||ana;Italian) | 3.3 (0.001) |
| Ju|'hoan_North | 46 | $2.3 \times 10^{-4} \pm 8.3 \times 10^{-6}$ (Ju|'hoan_South;Mandenka) | $4.1 \times 10^{-4} \pm 1.5 \times 10^{-5}$ (Ju|'hoan_South;Sardinian) | 10.3 ($< 1 \times 10^{-15}$) |
| Hai||om | 34 | $4.3 \times 10^{-4} \pm 1.6 \times 10^{-5}$ (Ju|'hoan_North;Yoruba) | $6.2 \times 10^{-4} \pm 2.5 \times 10^{-5}$ (Ju|'hoan_North;Sardinian) | 6.3 ($3 \times 10^{-10}$) |
| G|ui | 31 | $2.6 \times 10^{-4} \pm 1.3 \times 10^{-5}$ (Taa_North;Yoruba) | $4.3 \times 10^{-4} \pm 2.3 \times 10^{-5}$ (Taa_North;Sardinian) | 6.5 ($8 \times 10^{-11}$) |
| G||ana | 39 | $3.0 \times 10^{-4} \pm 4.3 \times 10^{-5}$ (G|ui;Yoruba) | $4.1 \times 10^{-4} \pm 2.5 \times 10^{-5}$ (Taa_West;Italian) | 2.3 (0.02) |
| G||ana | 4 | $1.4 \times 10^{-4} \pm 3.5 \times 10^{-5}$ (G|ui;Yoruba) | $1.1 \times 10^{-4} \pm 3.0 \times 10^{-5}$ (G|ui;Kalash) | 0.5 (0.6) |
| !Xuun | 43 | $3.2 \times 10^{-4} \pm 2.2 \times 10^{-5}$ (Ju|'hoan_South;Yoruba) | $4.8 \times 10^{-4} \pm 2.2 \times 10^{-5}$ (Ju|'hoan_North;Sardinian) | 5.1 ($3 \times 10^{-7}$) |
| !Xuun | 4 | $8.3 \times 10^{-5} \pm 1.4 \times 10^{-5}$ (Ju|'hoan_North;Yoruba) | $8.0 \times 10^{-5} \pm 1.4 \times 10^{-5}$ (Ju|'hoan_North;Druze) | 0.11 (0.9) |

Table 3: **Amplitudes of fitted admixture models for all southern African populations.** For each admixture event shown in Figure 4 in the main text in southern African populations, we show the amplitudes of the best "west African" populations (we include all Niger-Congo-speaking agriculturalist populations here, regardless of their geographic location) and "Eurasian" populations, and show the Z-score and corresponding P-value for a difference between the two.

13

| Population | One-admixture model date (gen. before present) | One-admixture model Z-score (P-value) | Two-admixture model dates (gen. before present) | Two-admixture model Z-scores (P-values) |
|---|---|---|---|---|
| G\|ui | 31 | 11.9 ($6 \times 10^{-33}$) | 109, 24 | 2.2 (0.014), 6.1 ($5 \times 10^{-10}$) |
| Hai\|\|om | 34 | 15.9 ($3 \times 10^{-57}$) | 51, 13 | 1.8 (0.035), 0.55 (0.29) |
| Ju\|'hoan_North | 46 | 18.8 ($4 \times 10^{-79}$) | 140, 38 | 1.7 (0.04), 8.4 ($2 \times 10^{-17}$) |
| Naro | 37 | 11.4 ($2 \times 10^{-30}$) | 43, 6 | 8.3 ($5 \times 10^{-17}$), 1.8 (0.04) |
| Shua | 32 | 14.0 ($8 \times 10^{-45}$) | 55, 18 | 1.9 (0.03), 1.8 (0.04) |
| Taa_North | 30 | 10.4 ($1 \times 10^{-25}$) | 44, 6 | 8.1 ($3 \times 10^{-16}$), 1.9 (0.03) |
| Tshwa | 23 | 10.1 ($3 \times 10^{-24}$) | 45, 6 | 8.0 ($6 \times 10^{-16}$), 2.0 (0.02) |
| ǂHoan | 14 | 7.1 ($6 \times 10^{-13}$) | 32, 6 | 3.1 (0.001), 1.8 (0.04) |

Table 4: **Southern African populations with a single inferred admixture event.** For each southern African population where Figure 4 in the main text shows a single admixture event, we show the admixture times inferred from both the one- and two-admixture models. Additionally shown are the Z-scores and P-values for each admixture time (we used a P-value threshold of 0.01 to call an admixture event as "significant", so for all of these populations at least one mixture time in the two-admixture model is non-significant). The G|ui and Ju|'hoan_North stand out as having borderline evidence for an old admixture event around 100 generations ago.

| Target Population | Reference populations | $f_3$ | Z-score |
|---|---|---|---|
| Afar | Sardinian, Sudanese | -0.026 | -66.4 |
| Afar | Sardinian, Anuak | -0.026 | -63.8 |
| Afar | French Basque, Sudanese | -0.025 | -61.0 |
| Amhara | Tuscan, Sudanese | -0.028 | -91.4 |
| Amhara | Tuscan, Anuak | -0.028 | -90.7 |
| Amhara | Samaritians, Anuak | -0.027 | -85.8 |
| Anuak | Ari Blacksmith, Sudanese | -0.001 | -8.2 |
| Anuak | Ari Cultivator, Sudanese | -0.001 | -8.6 |
| Anuak | Wolayta, Sudanese | -0.001 | -5.5 |
| Ari Cultivator | Sardinian, Ju|'hoan | -0.013 | -25.7 |
| Ari Cultivator | Samaritian, Ju|'hoan | -0.013 | -21.1 |
| Ari Cultivator | Tuscan, Ju|'hoan | -0.013 | -24.1 |
| Ethiopian Somali | Sardinian, Sudanese | -0.024 | -64.2 |
| Ethiopian Somali | Sardinian, Anuak | -0.024 | -64.1 |
| Ethiopian Somali | Cypriot, Sudanese | -0.023 | -64.2 |
| Luhya | Sardinian, Biaka | -0.005 | -16.8 |
| Luhya | Bedouin, Biaka | -0.005 | -17.4 |
| Luhya | Yemenite Jews, Biaka | -0.005 | -17.4 |
| Maasai | Sardinian, Mbuti | -0.021 | -54.5 |
| Maasai | Cypriot, Mbuti | -0.021 | -54.6 |
| Maasai | Samaritian, Mbuti | -0.021 | -42.8 |
| Oromo | Sardinian, Sudanese | -0.031 | -95.7 |
| Oromo | Sardinian, Anuak | -0.030 | -93.2 |
| Oromo | Samaritian, Anuak | -0.030 | -70.2 |
| Somali | Sardinian, Sudanese | -0.022 | -60.3 |
| Somali | Sardinian, Anuak | -0.022 | -60.5 |
| Somali | French Basque, Sudanese | -0.021 | -55.3 |
| Tygray | Sardinian, Sudanese | -0.029 | -88.7 |
| Tygray | Sardinian, Anuak | -0.029 | -87.2 |
| Tygray | Cypriot, Sudanese | -0.028 | -88.2 |
| Wolayta | Sardinian, Gumuz | -0.025 | -69.6 |
| Wolayta | Cypriot, Gumuz | -0.024 | -69.3 |
| Wolayta | Yemenite Jews, Gumuz | -0.024 | -72.0 |

Table 5: **Three-population tests for treeness in eastern Africa.** We performed three-population tests on all eastern African populations from Pagani et al. [2012] and the HapMap 3. For each population with at least one $f_3$ statistic with a Z-score less than $-3$, we show details of the three smallest $f_3$ statistics: the names of the reference populations, the value of the statistic, and the Z-score. A Z-score of less than -3 corresponds to a p-value of less than 0.001. The eastern African populations with no significantly negative $f_3$ statistics are the Sudanese, Gumuz, and Ari Blacksmith

| Population | Mixture date in generations | Z-score (Eurasian versus non-Eurasian) | Z-score (west African versus non-west African) |
|---|---|---|---|
| Wolayta | 59 | 6.5 | 1.0 |
| Tygray | 88 | 12.7 | 0.9 |
| Somali | 109 | 8.6 | 0.7 |
| Sandawe | 130 | 7.0 | 2.4 |
| Sandawe | 130 | 7.0 | 2.4 |
| Sandawe | 27 | 1.2 | 0.2 |
| Oromo | 95 | 12.9 | 1.0 |
| Oromo | 8 | 2.2 | 0.6 |
| Maasai | 88 | 22.5 | 1.1 |
| Maasai | 8 | 3.1 | 0.1 |
| Luhya | 106 | 4.6 | 0.8 |
| Luhya | 16 | 1.0 | 0.3 |
| Hadza | 61 | 3.2 | 0.5 |
| Gumuz | 5 | 5.1 | 0.6 |
| Gumuz | 112 | 2.3 | 0.6 |
| ESomali | 107 | 9.3 | 0.6 |
| AriCultivator | 104 | 7.2 | 1.5 |
| AriBlacksmith | 109 | 6.9 | 1.2 |
| Anuanak | 82 | 0.9 | 0.4 |
| Amhara | 115 | 14.2 | 1.5 |
| Amhara | 11 | 2.3 | 0.8 |
| Afar | 81 | 7.7 | 0.9 |

Table 6: **Amplitudes of fitted admixture models for all eastern African populations.** We split the reference populations used to calculate weighted LD in each population into "west African" (including all Niger-Congo-speaking agriculturalist populations), "Eurasian", and "other African" (including populations from southern and eastern Africa). We computed a Z-score comparing the largest amplitude where one reference population is Eurasian to the largest amplitude where neither population is Eurasian, as well as a Z-score comparing the largest amplitude where one reference population is west African to the largest amplitude where neither population is west African. Highlighted in grey is the comparison involving the overall maximum amplitude and thus the source population reported in Figure 4.

| Population | # individuals |
|---|---|
| AMHARA | 26 |
| Khwe | 17 |
| Adygei | 17 |
| AFAR | 12 |
| ANUAK | 23 |
| ARIBLACKSMITH | 17 |
| ARICULTIVATOR | 24 |
| Armenians | 19 |
| BantuKenya | 11 |
| BantuSouthAfrica | 8 |
| Basque | 24 |
| Bedouin | 45 |
| BiakaPygmy | 22 |
| CEU (Utah) | 112 |
| Cypriots | 12 |
| Druze | 42 |
| Egyptians | 12 |
| ESOMALI | 17 |
| Ethiopians | 19 |
| Han | 34 |
| Japanese | 28 |
| French | 28 |
| Georgians | 20 |
| GuiGhanaKgal | 15 |
| GUMUZ | 19 |
| HADZA | 17 |
| Hungarians | 20 |
| Iranians | 20 |
| Italian | 12 |
| Jordanians | 20 |
| Juhoansi | 23 |
| Kalash | 23 |
| Khomani | 39 |
| Lebanese | 8 |
| LWK (Luhya) | 90 |
| Mandenka | 22 |
| MbutiPygmy | 13 |
| MKK (Maasai) | 143 |
| Moroccans | 10 |
| Mozabite | 27 |
| Nama | 20 |
| Orcadian | 15 |
| OROMO | 21 |
| Palestinian | 46 |
| Russian | 25 |
| SANDAWE | 28 |
| Sardinian | 28 |
| SOMALI | 23 |
| SUDANESE | 24 |
| Syrians | 16 |
| TSI (Tuscany) | 88 |
| TYGRAY | 21 |
| WOLAYTA | 8 |
| Xun | 19 |
| YRI (Yoruba) | 113 |

Table 7: **Populations typed on an Illumina array and used in analyses of multiple mixture events.** Labels are taken from the papers in which the samples were first reported.

17

# 3 Supplementary Figures

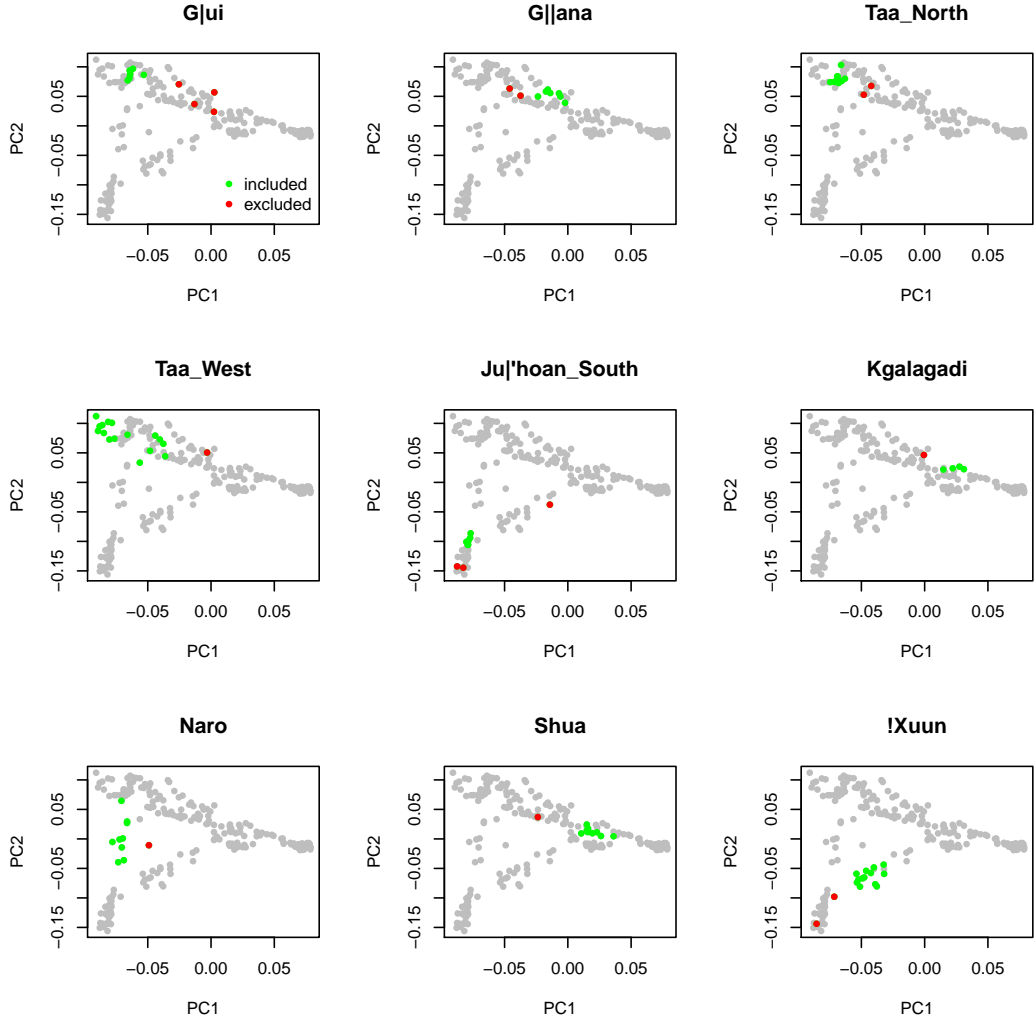

Figure 1: **Outliers in southern African data.** We ran smartpca [Patterson et al., 2006] on the southern African samples, and visually examined the PCA plots for individuals that appeared to be outliers with respect to other individuals with the same population label. For this analysis, following Pickrell et al. [2012], we used only the SNPs ascertained in a single Ju|'hoan (HGDP "San") individual in order to expose the population structure within the southern Africa Khoisan. Shown in each panel are all the individuals removed from analysis (red circles), along with the other individuals from their population (green circles).

19

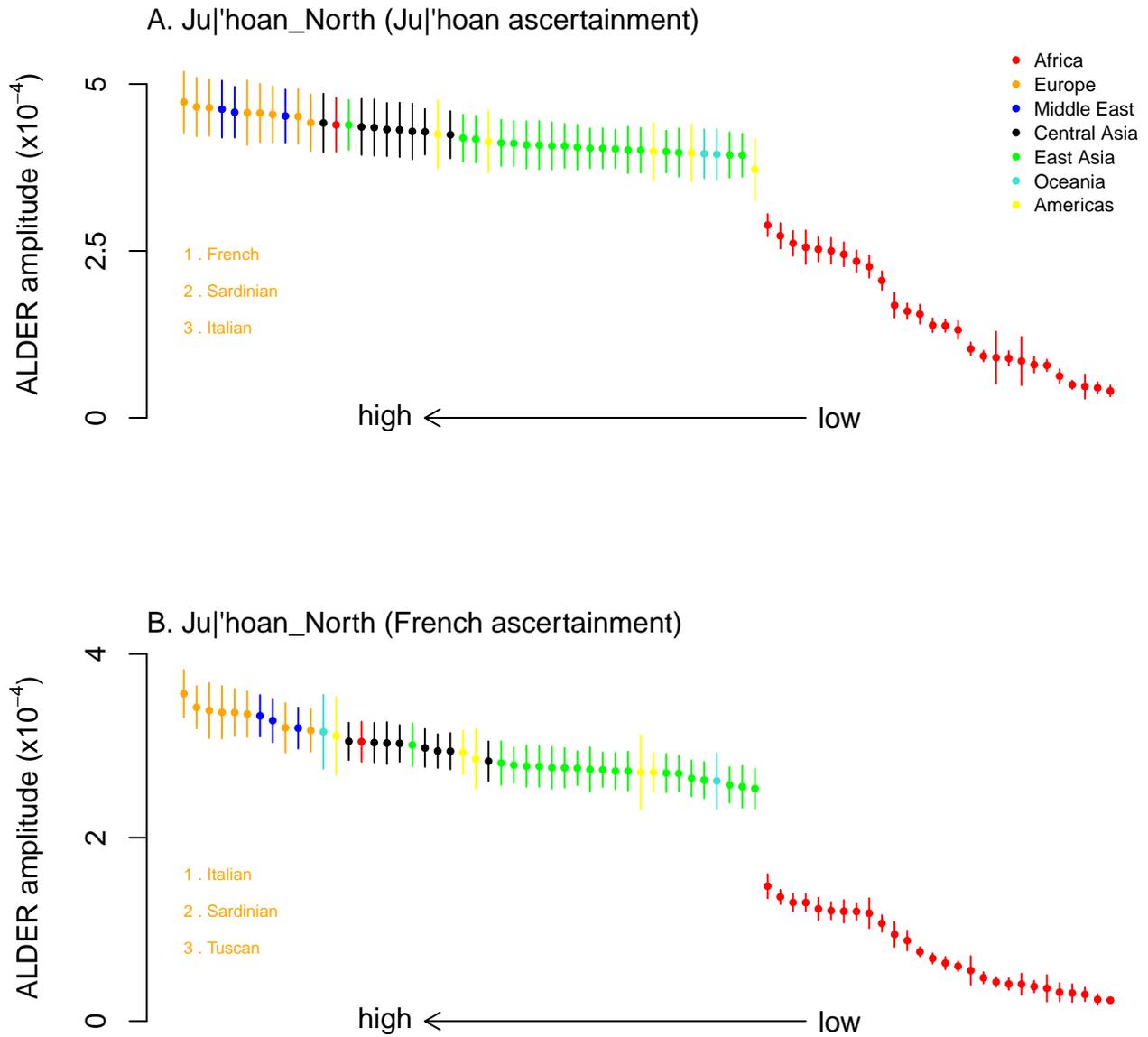

Figure 2: **Signal of west Eurasian ancestry in the Ju|'hoan_North is robust to SNP ascertainment.** This figure is identical to Figure 1 in the main text, except the amplitudes were calculated using only SNPs on the Human Origins Array from individual ascertainment panels.

20

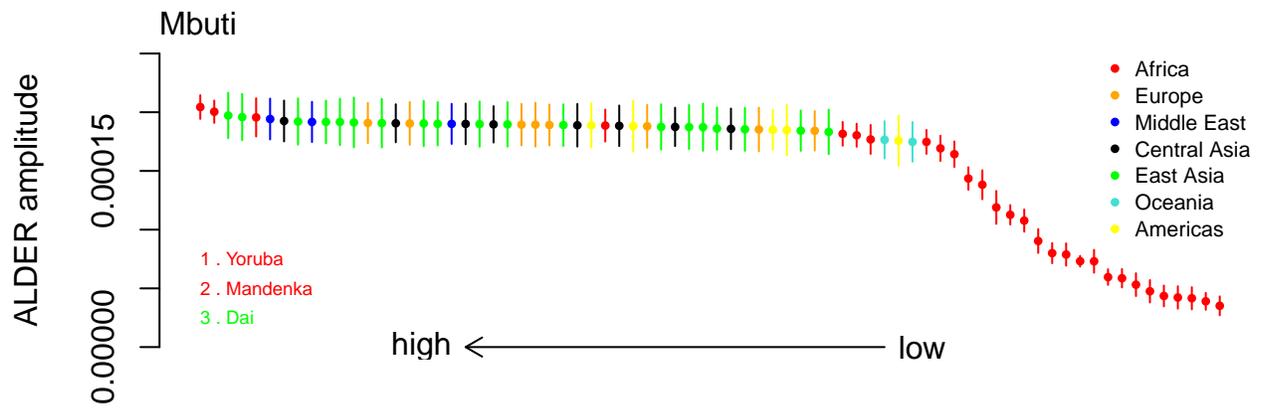

Figure 3: **Identifying sources of admixture in the Mbuti.** We calculated weighted LD curves in the Mbuti, using the Mbuti themselves as one reference population and a set of other worldwide populations as the other reference. As in Figure 1 in the main text, we show the estimated amplitudes of these LD curves, colored according to the continent of the reference population.

21

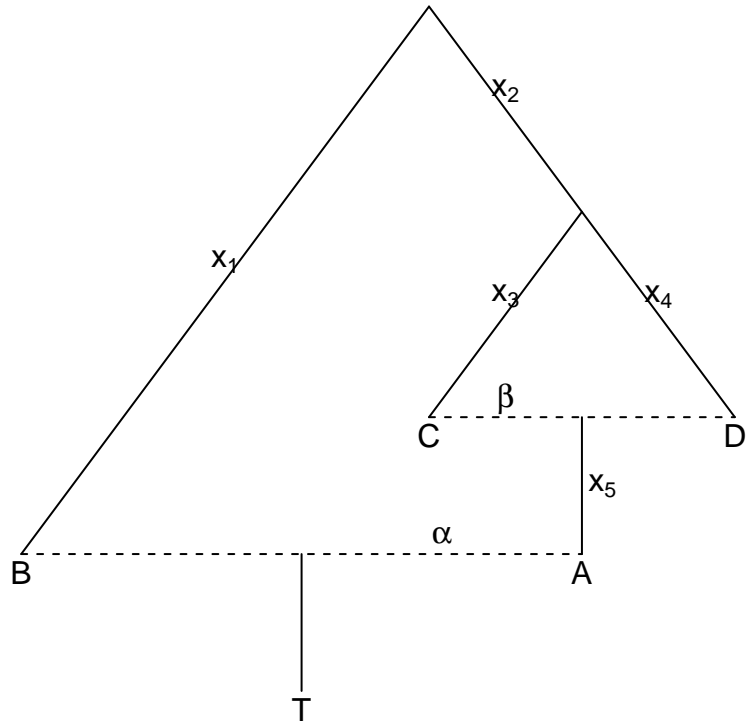

Figure 4: **Schematic of a history with two admixture events.** Shown is an example admixture graph, where solid lines represent population relationships and dotted lines represent admixture events. The capital letters represent populations, $\alpha$ and $\beta$ represent the admixture proportions in the two mixture events, and the $x$ parameters represent branch lengths in units of genetic drift.

22

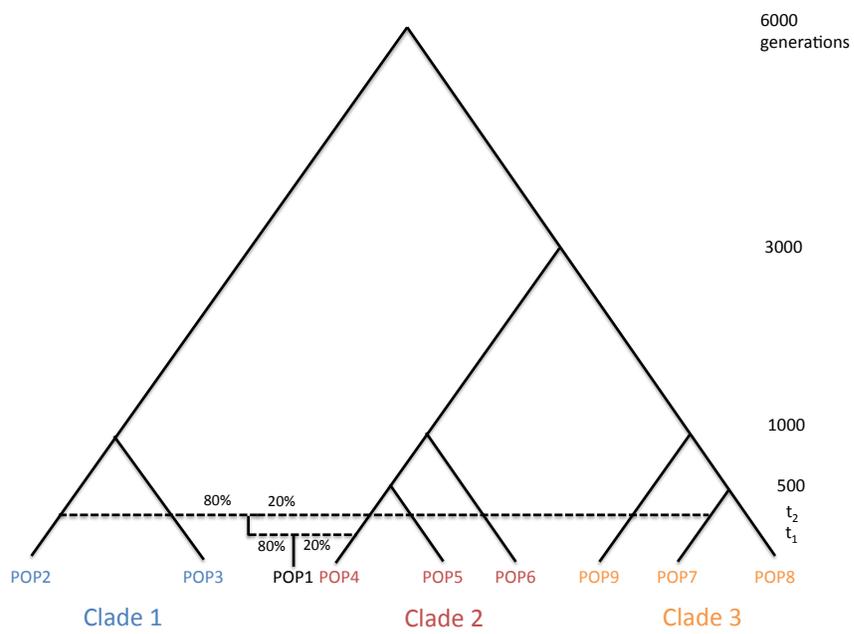

Figure 5: **Population history used in simulations.**

23

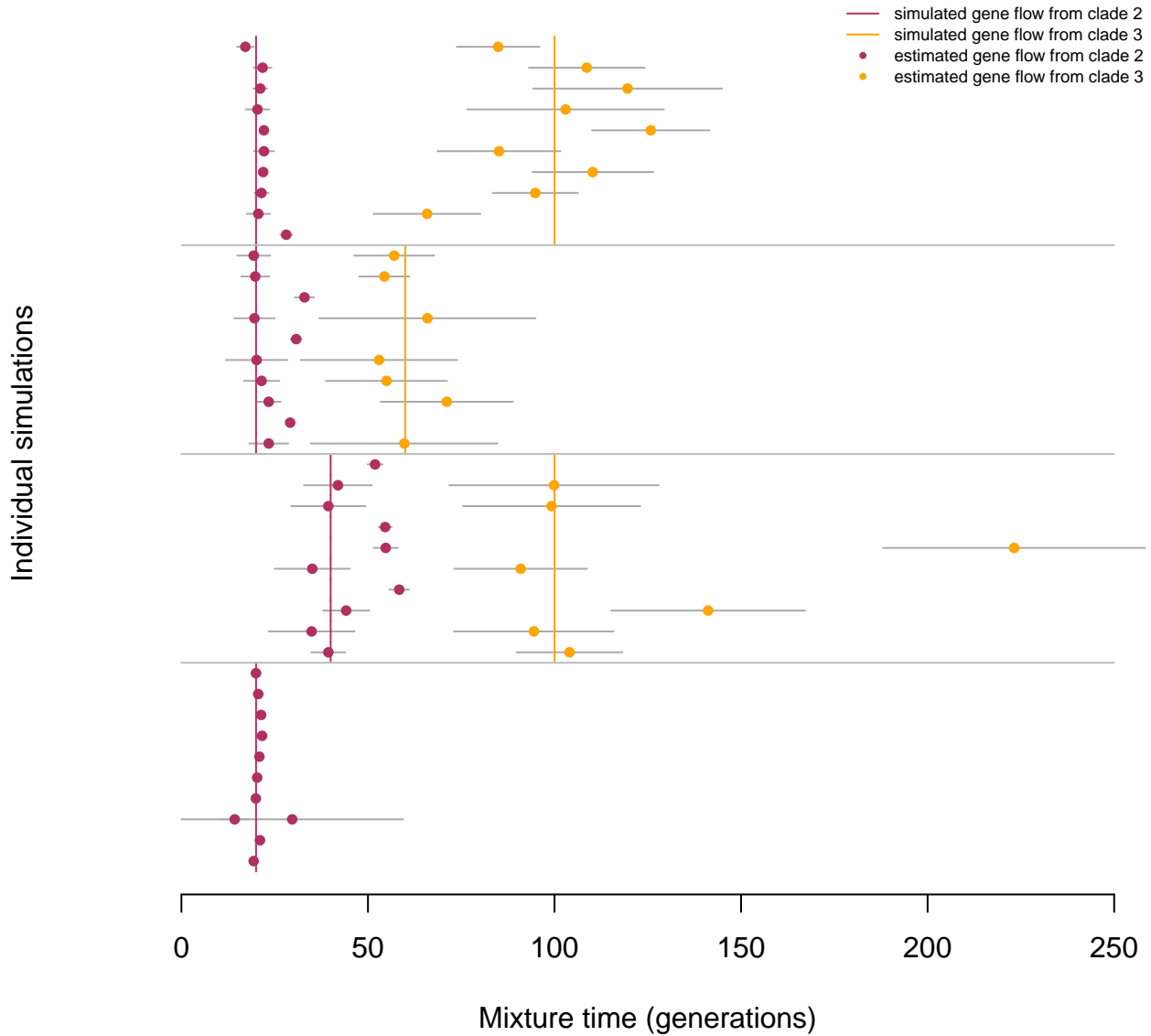

Figure 6: **Simulation results under pulse admixture model.** We tested our method for estimating multiple mixture dates using simulations. In each simulation, we generated data from the demographic model in Supplementary Figure 5 with mixture dates denoted by the colored lines. We then ran our method; each point represents an estimated mixture event, and shown are the within-simulation standard errors. Points are colored according to the inferred source of the admixture. In no simulation did we detect more than two admixture events.

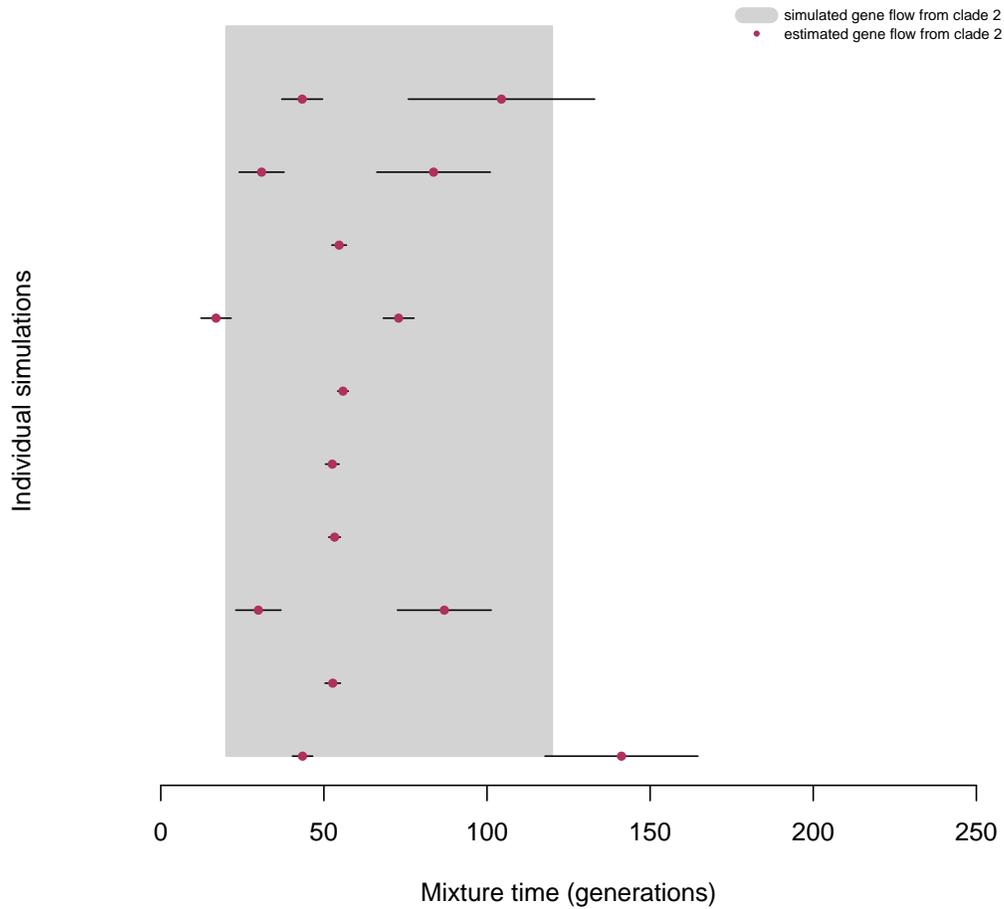

Figure 7: **Simulation results under continuous gene flow model.** We tested our method for estimating multiple mixture dates using simulations. In each simulation, we generated data from the demographic model in Supplementary Figure 5 with only gene flow from clade 2. In these simulations, instead of 20% admixture at a single point in time, we simulated 0.2% admixture per generation over 100 generations (for a total of 20% admixture). The grey box shows the 100 generations over which admixture is occurring. We then ran our method; each point represents an inferred mixture event, and shown are the within-simulation standard errors.

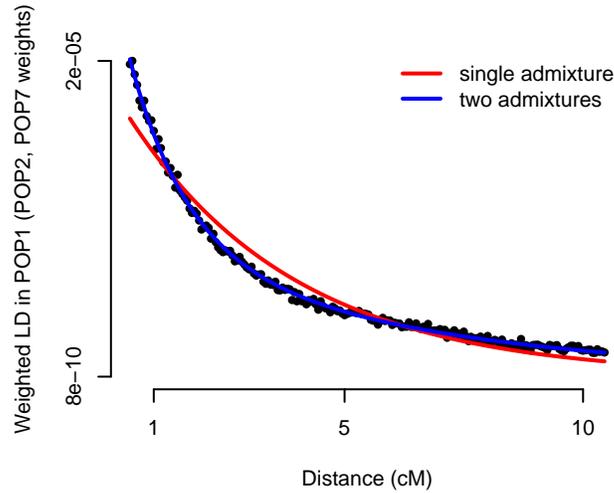

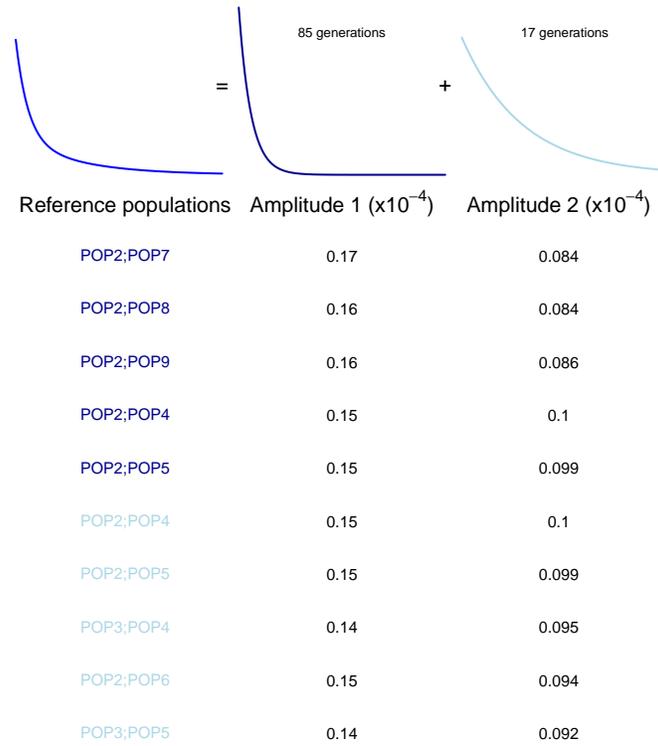

Figure 8: **Example simulation with one recent and one old admixture event.** We simulated data under the demographic model in Supplementary Figure 5, where $t_1 = 20$ and $t_2 = 100$. We then fit a model of multiple mixture events. Shown in red is the fitted model with a single admixture event; in blue is the fitted model with two admixture events. Below the graph are the five population pairs with the largest weights on the first and second inferred admixture events. In both cases the inferred mixing population is the correct one.

26

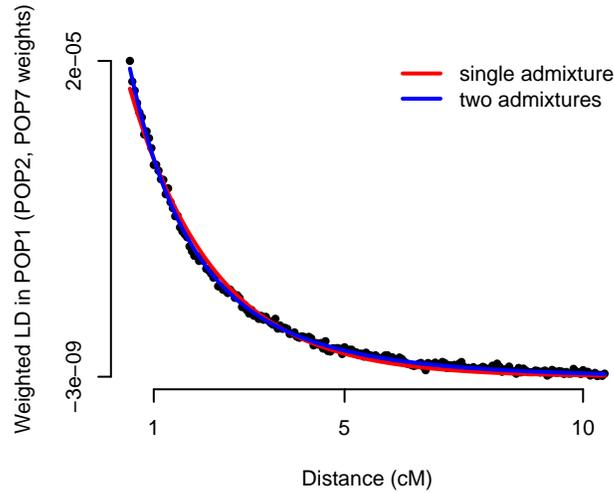

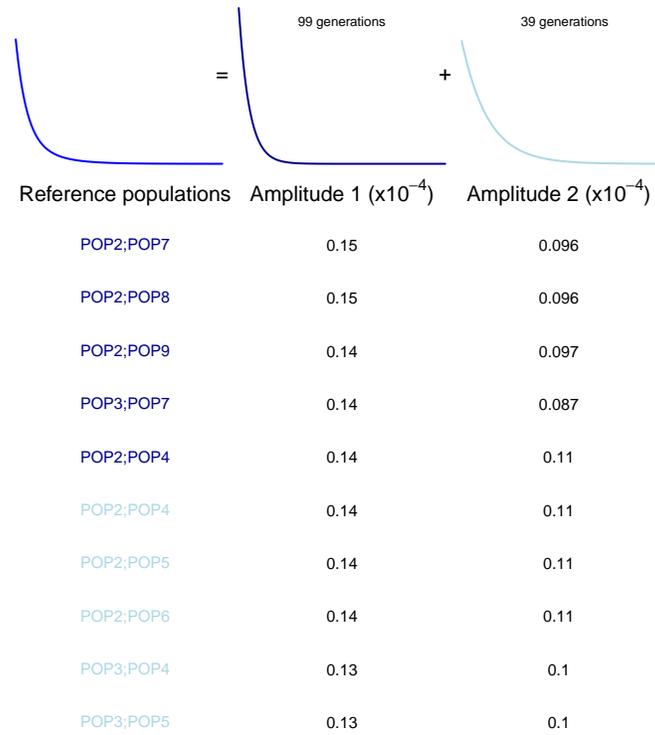

Figure 9: **Example simulation with two old admixture events.** We simulated data under the demographic model in Supplementary Figure 5, where $t_1 = 40$ and $t_2 = 100$. We then fit a model of multiple mixture events. Shown in red is the fitted model with a single admixture event; in blue is the fitted model with two admixture events. Below the graph are the five population pairs with the largest weights on the first and second inferred admixture events. In both cases the inferred mixing population is the correct one.

27

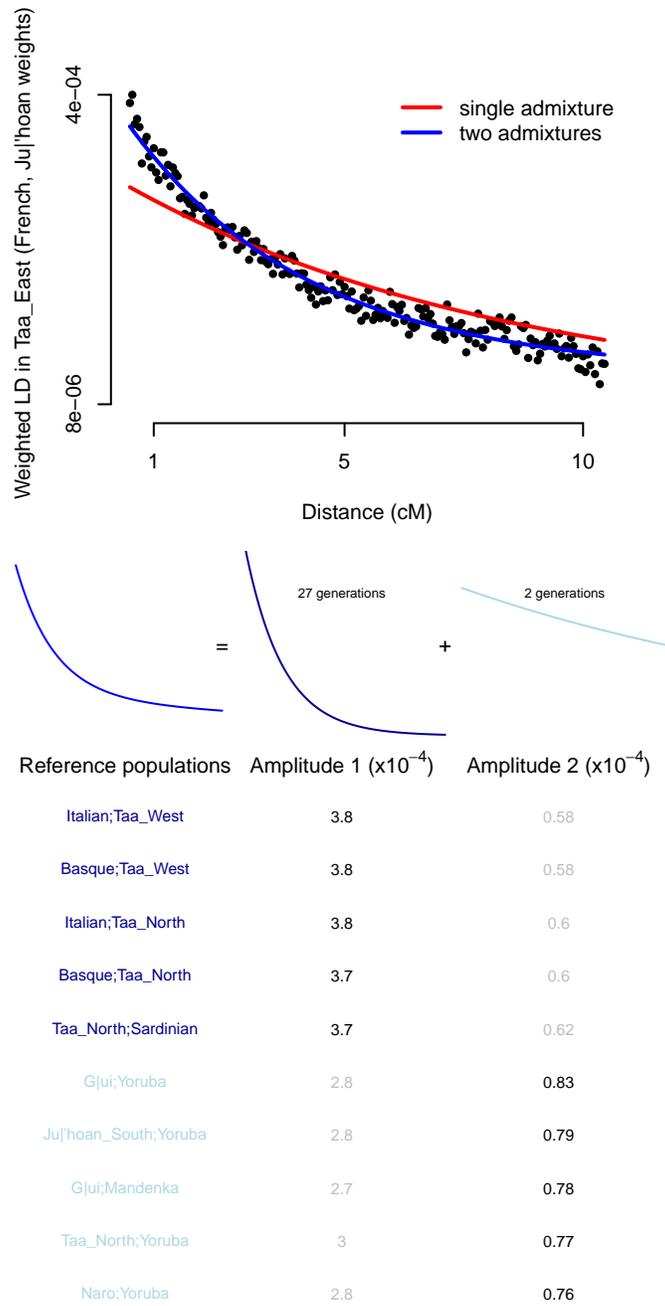

Figure 10: **Fitted admixture model in the Taa_East.** See the caption to Figure 3 in the main text for details.

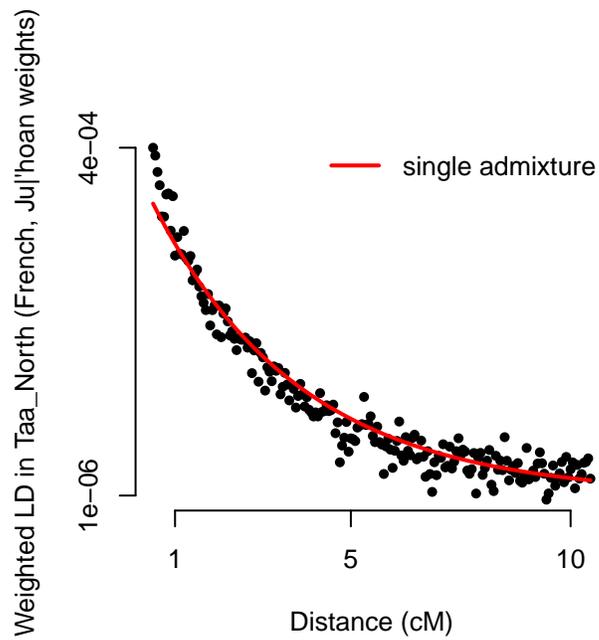

| Reference populations | Amplitude 1 (x$10^{-4}$) |
|---|---|
| Druze;Taa_West | 4 |
| Druze;G|ui | 3.9 |
| Italian;Taa_West | 3.9 |
| G|ui;Italian | 3.9 |
| French;Taa_West | 3.9 |
| Sardinian;Taa_West | 3.9 |
| G|ui;Sardinian | 3.9 |
| French;G|ui | 3.9 |
| Druze;Naro | 3.9 |
| Tuscan;Taa_West | 3.8 |

Figure 11: **Fitted admixture model in the Taa_North.** See the caption to Figure 3 in the main text for details.

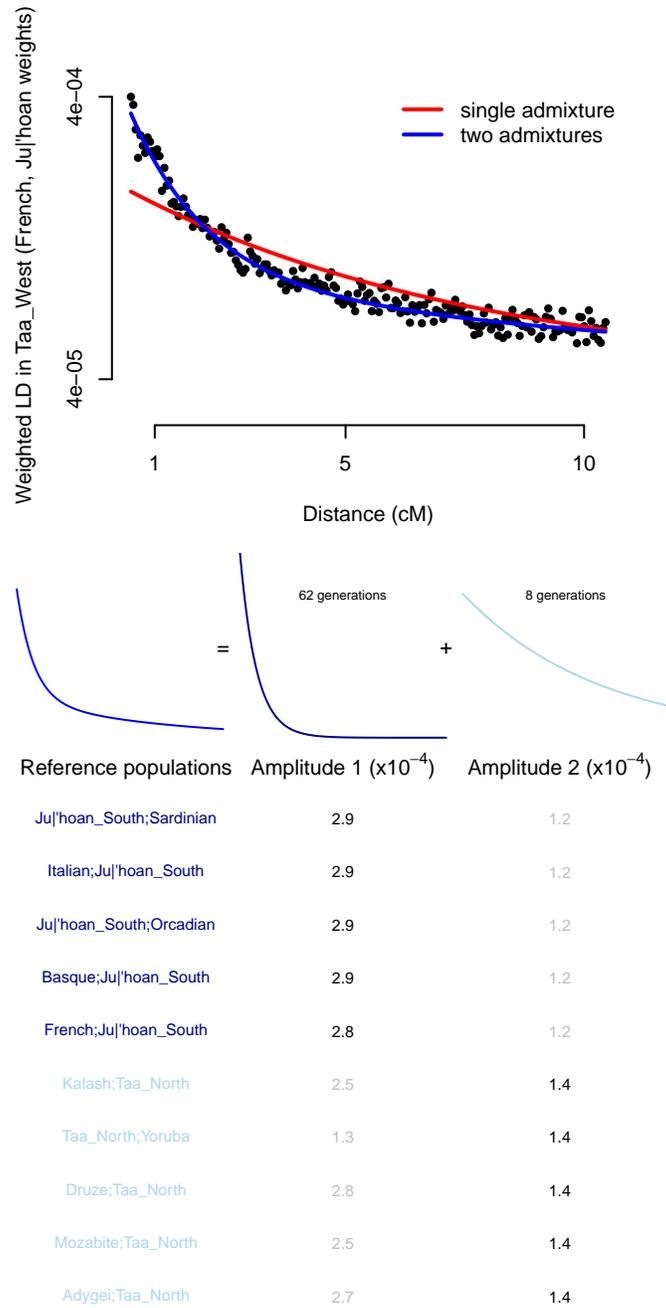

Figure 12: **Fitted admixture model in the Taa_West.** See the caption to Figure 3 in the main text for details.

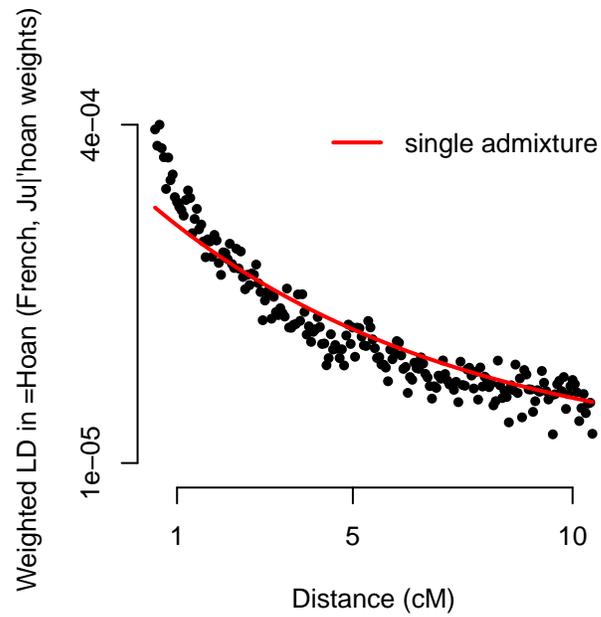

| Reference populations | Amplitude 1 (x$10^{-4}$) |
|---|---|
| Druze;Taa_North | 3.5 |
| Taa_North;Sardinian | 3.4 |
| Taa_North;Orcadian | 3.4 |
| Italian;Taa_North | 3.4 |
| Adygei;Taa_North | 3.4 |
| Bedouin;Taa_North | 3.4 |
| Taa_North;Tuscan | 3.4 |
| Basque;Taa_North | 3.4 |
| Mozabite;Taa_North | 3.4 |
| French;Taa_North | 3.4 |

Figure 13: **Fitted admixture model in the ǂHoan.** See the caption to Figure 3 in the main text for details.

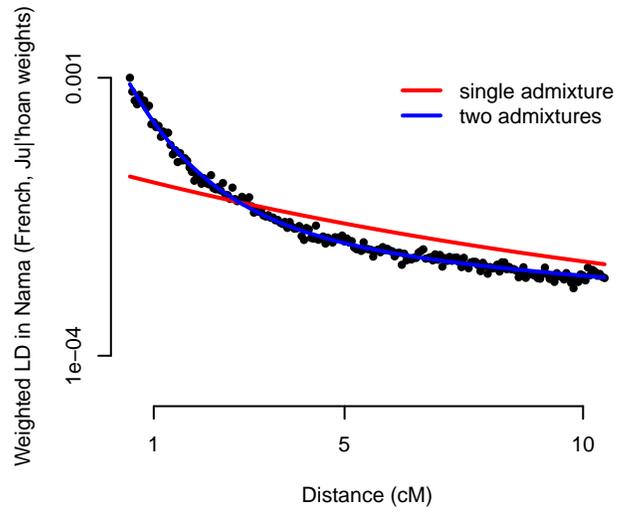

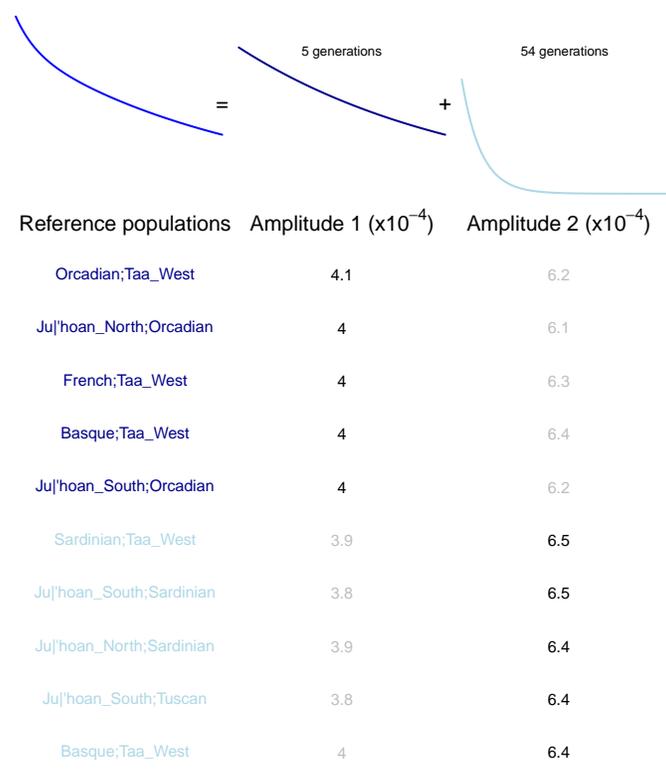

| Reference populations | Amplitude 1 (x$10^{-4}$) | Amplitude 2 (x$10^{-4}$) |
|---|---|---|
| Orcadian;Taa_West | 4.1 | 6.2 |
| Ju|'hoan_North;Orcadian | 4 | 6.1 |
| French;Taa_West | 4 | 6.3 |
| Basque;Taa_West | 4 | 6.4 |
| Ju|'hoan_South;Orcadian | 4 | 6.2 |
| Sardinian;Taa_West | 3.9 | 6.5 |
| Ju|'hoan_South;Sardinian | 3.8 | 6.5 |
| Ju|'hoan_North;Sardinian | 3.9 | 6.4 |
| Ju|'hoan_South;Tuscan | 3.8 | 6.4 |
| Basque;Taa_West | 4 | 6.4 |

Figure 14: **Fitted admixture model in the Nama.** See the caption to Figure 3 in the main text for details.

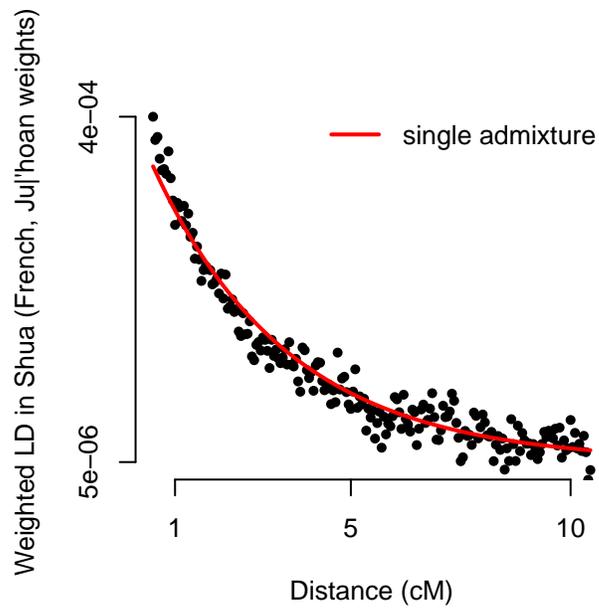

| Reference populations | Amplitude 1 (x$10^{-4}$) |
|---|---|
| Druze;Ju\|'hoan_South | 4.5 |
| Ju\|'hoan_South;Sardinian | 4.5 |
| Druze;Ju\|'hoan_North | 4.5 |
| Ju\|'hoan_North;Sardinian | 4.5 |
| Ju\|'hoan_South;Tuscan | 4.5 |
| Italian;Ju\|'hoan_South | 4.5 |
| Ju\|'hoan_South;Russian | 4.5 |
| Ju\|'hoan_North;Tuscan | 4.5 |
| Italian;Ju\|'hoan_North | 4.5 |
| Ju\|'hoan_North;Russian | 4.5 |

Figure 15: **Fitted admixture model in the Shua.** See the caption to Figure 3 in the main text for details.

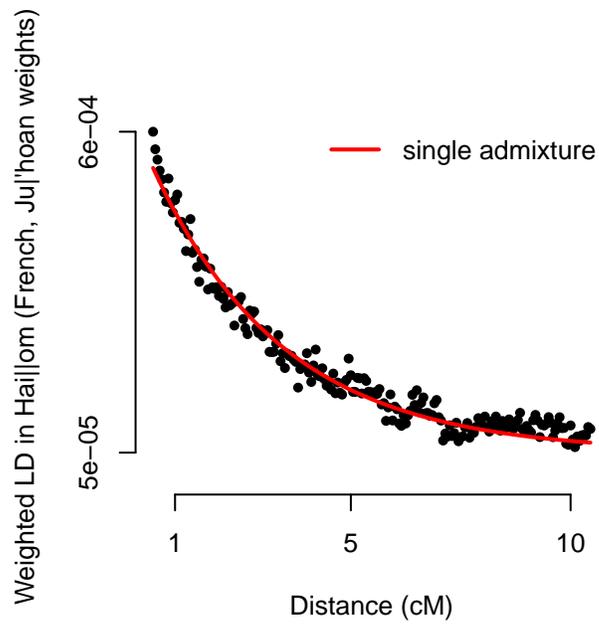

| Reference populations | Amplitude 1 (x$10^{-4}$) |
|---|---|
| Ju|'hoan_North;Sardinian | 6.2 |
| Italian;Ju|'hoan_North | 6.1 |
| French;Ju|'hoan_North | 6 |
| Ju|'hoan_North;Tuscan | 6 |
| Druze;Ju|'hoan_North | 6 |
| Basque;Ju|'hoan_North | 6 |
| Ju|'hoan_South;Sardinian | 6 |
| Ju|'hoan_North;Palestinian | 5.9 |
| Adygei;Ju|'hoan_North | 5.9 |
| Ju|'hoan_North;Orcadian | 5.9 |

Figure 16: **Fitted admixture model in the Hai||om.** See the caption to Figure 3 in the main text for details.

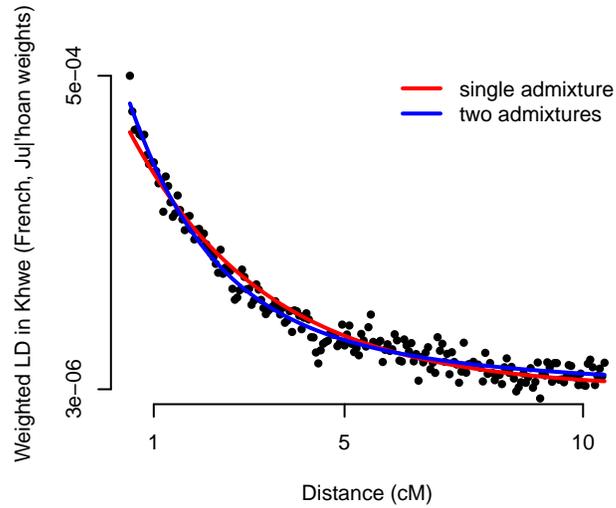

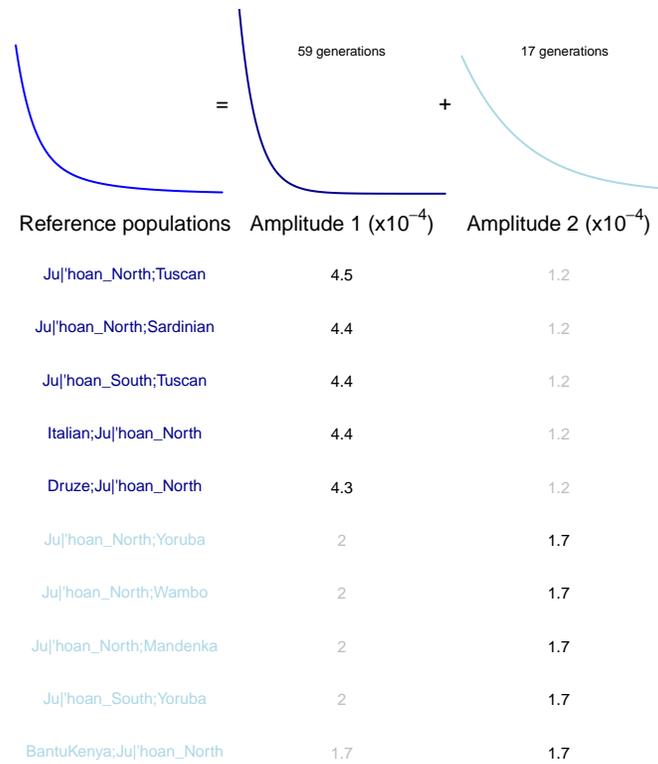

| Reference populations | Amplitude 1 (x$10^{-4}$) | Amplitude 2 (x$10^{-4}$) |
|---|---|---|
| Ju|'hoan_North;Tuscan | 4.5 | 1.2 |
| Ju|'hoan_North;Sardinian | 4.4 | 1.2 |
| Ju|'hoan_South;Tuscan | 4.4 | 1.2 |
| Italian;Ju|'hoan_North | 4.4 | 1.2 |
| Druze;Ju|'hoan_North | 4.3 | 1.2 |
| Ju|'hoan_North;Yoruba | 2 | 1.7 |
| Ju|'hoan_North;Wambo | 2 | 1.7 |
| Ju|'hoan_North;Mandenka | 2 | 1.7 |
| Ju|'hoan_South;Yoruba | 2 | 1.7 |
| BantuKenya;Ju|'hoan_North | 1.7 | 1.7 |

Figure 17: **Fitted admixture model in the Khwe.** See the caption to Figure 3 in the main text for details.

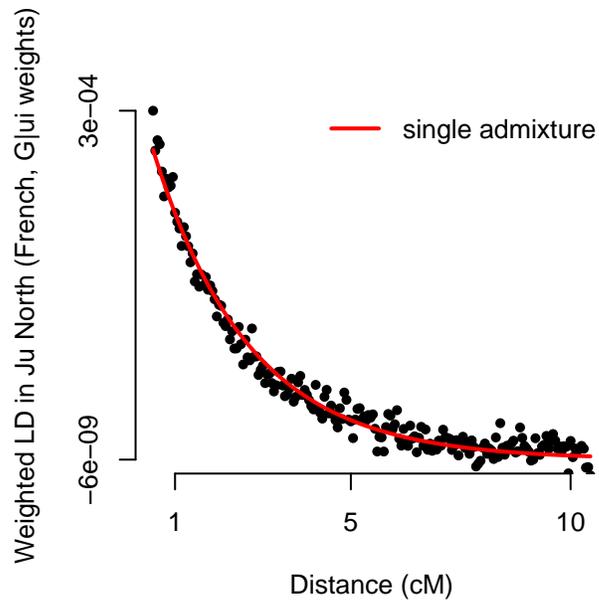

| Reference populations | Amplitude 1 (x10$^{-4}$) |
|---|---|
| Ju|'hoan_South;Sardinian | 4.1 |
| Ju|'hoan_South;Orcadian | 4.1 |
| Italian;Ju|'hoan_South | 4.1 |
| Basque;Ju|'hoan_South | 4.1 |
| Ju|'hoan_South;Tuscan | 4.1 |
| French;Ju|'hoan_South | 4.1 |
| Druze;Ju|'hoan_South | 4.1 |
| Adygei;Ju|'hoan_South | 4 |
| Bedouin;Ju|'hoan_South | 4 |
| Ju|'hoan_South;Russian | 4 |

Figure 18: **Fitted admixture model in the Ju|'hoan North.** See the caption to Figure 3 in the main text for details.

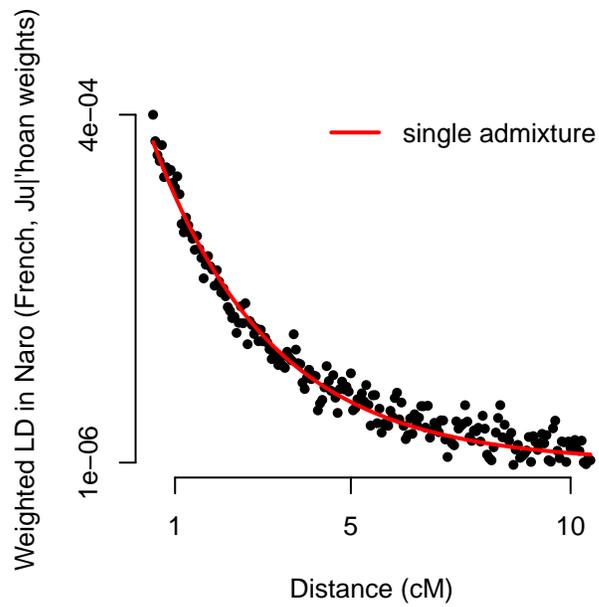

| Reference populations | Amplitude 1 (x$10^{-4}$) |
|---|---|
| Ju\|'hoan_South;Sardinian | 4.9 |
| Italian;Ju\|'hoan_South | 4.8 |
| Druze;Ju\|'hoan_South | 4.8 |
| Ju\|'hoan_North;Sardinian | 4.8 |
| Ju\|'hoan_South;Tuscan | 4.8 |
| Basque;Ju\|'hoan_South | 4.8 |
| Italian;Ju\|'hoan_North | 4.7 |
| Bedouin;Ju\|'hoan_South | 4.7 |
| Druze;Ju\|'hoan_North | 4.7 |
| French;Ju\|'hoan_South | 4.7 |

Figure 19: **Fitted admixture model in the Naro.** See the caption to Figure 3 in the main text for details.

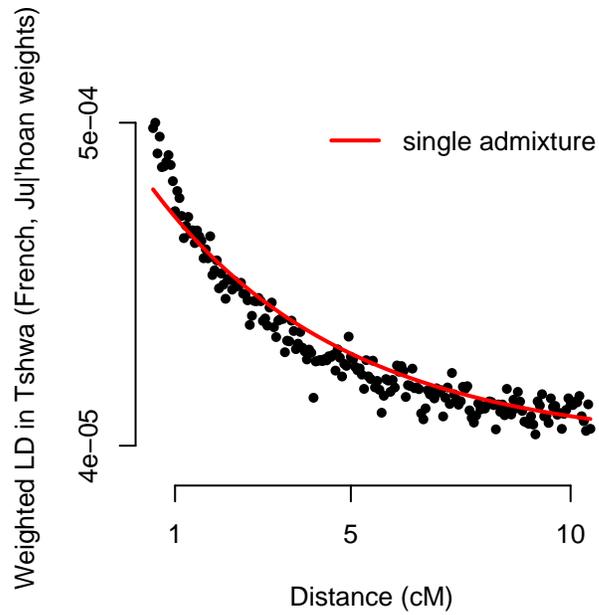

| Reference populations | Amplitude 1 (x$10^{-4}$) |
|---|---|
| Italian;Ju|'hoan_South | 4.2 |
| Ju|'hoan_South;Tuscan | 4.2 |
| Ju|'hoan_South;Sardinian | 4.2 |
| Basque;Ju|'hoan_South | 4.2 |
| Italian;Ju|'hoan_North | 4.2 |
| Ju|'hoan_North;Tuscan | 4.2 |
| Druze;Ju|'hoan_South | 4.2 |
| Ju|'hoan_North;Sardinian | 4.2 |
| French;Ju|'hoan_South | 4.2 |
| Ju|'hoan_South;Orcadian | 4.2 |

Figure 20: **Fitted admixture model in the Tshwa.** See the caption to Figure 3 in the main text for details.

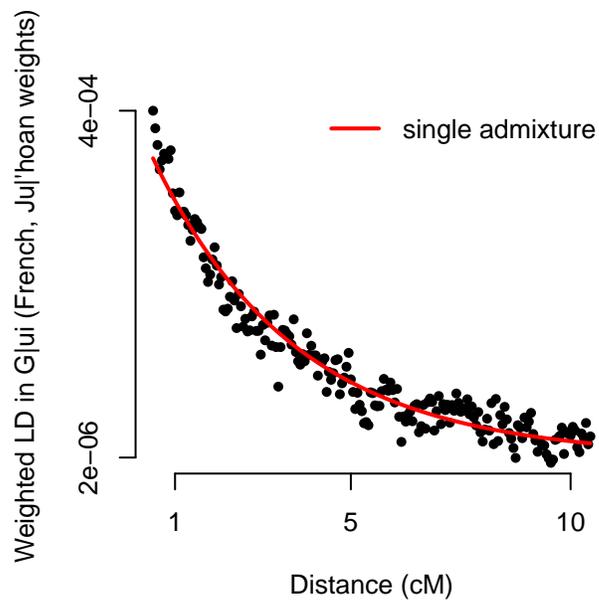

| Reference populations | Amplitude 1 (x$10^{-4}$) |
|---|---|
| Taa_North;Sardinian | 4.3 |
| Sardinian;Taa_West | 4.2 |
| Basque;Taa_North | 4.2 |
| Druze;Taa_North | 4.2 |
| Italian;Taa_North | 4.2 |
| Taa_North;Orcadian | 4.2 |
| Italian;Taa_West | 4.2 |
| Basque;Taa_West | 4.2 |
| Taa_North;Tuscan | 4.2 |
| French;Taa_North | 4.2 |

Figure 21: **Fitted admixture model in the G|ui.** See the caption to Figure 3 in the main text for details.

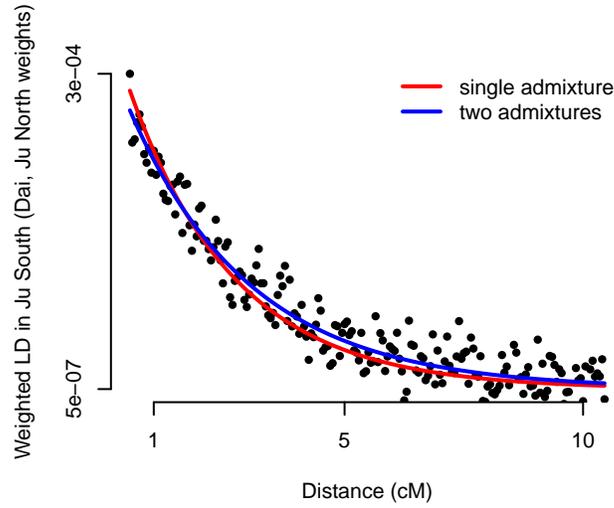

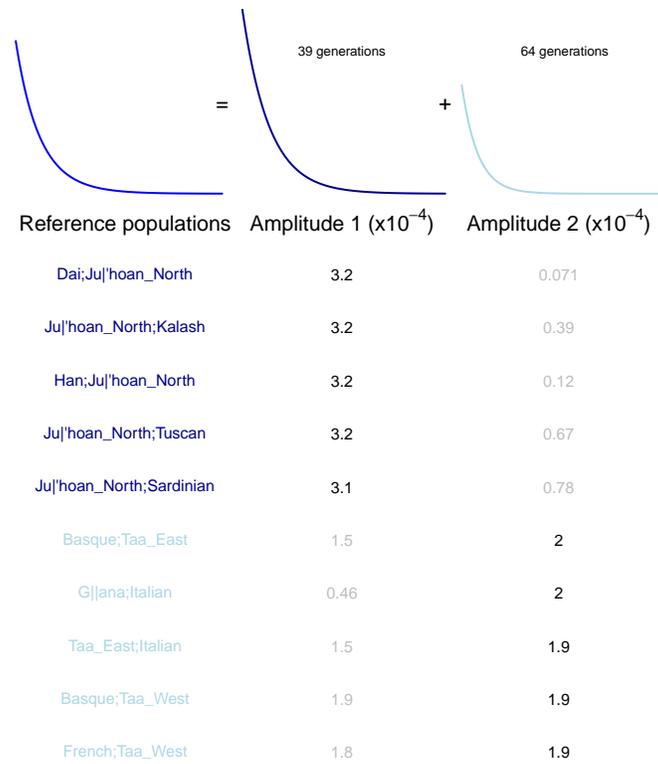

| Reference populations | Amplitude 1 (x$10^{-4}$) | Amplitude 2 (x$10^{-4}$) |
|---|---|---|
| Dai;Ju\|'hoan_North | 3.2 | 0.071 |
| Ju\|'hoan_North;Kalash | 3.2 | 0.39 |
| Han;Ju\|'hoan_North | 3.2 | 0.12 |
| Ju\|'hoan_North;Tuscan | 3.2 | 0.67 |
| Ju\|'hoan_North;Sardinian | 3.1 | 0.78 |
| Basque;Taa_East | 1.5 | 2 |
| G\|\|ana;Italian | 0.46 | 2 |
| Taa_East;Italian | 1.5 | 1.9 |
| Basque;Taa_West | 1.9 | 1.9 |
| French;Taa_West | 1.8 | 1.9 |

Figure 22: **Fitted admixture model in the Ju|'hoan_South.** See the caption to Figure 3 in the main text for details.

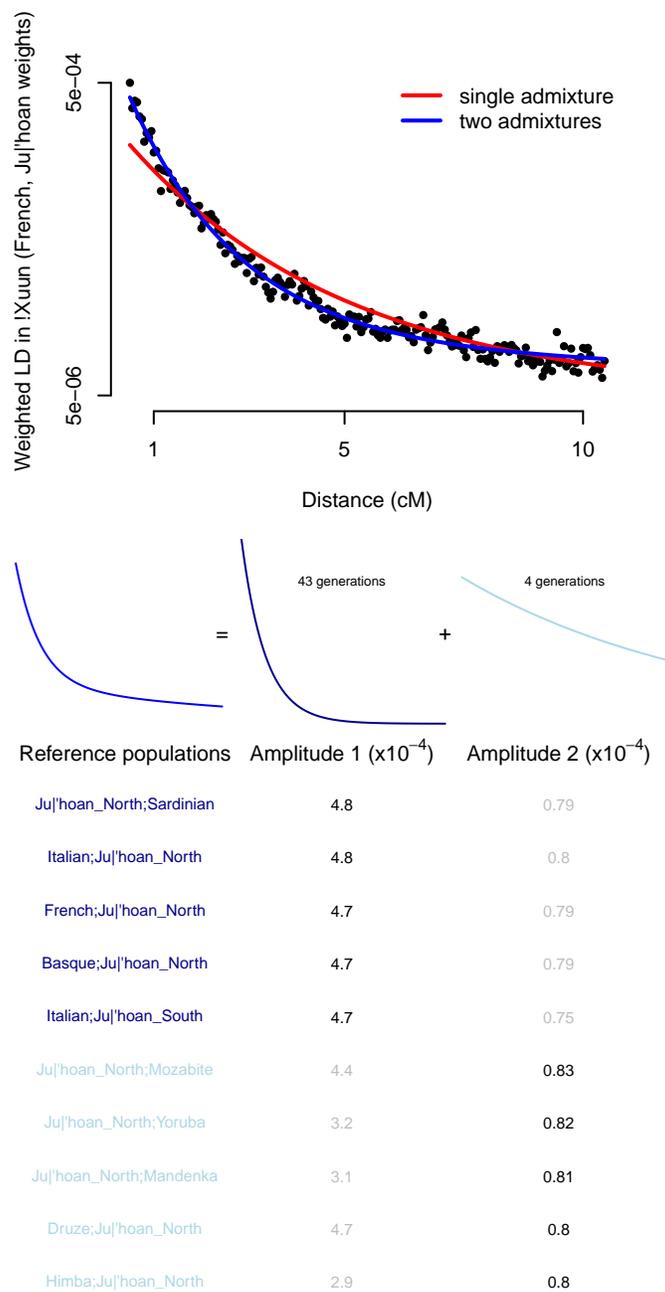

Figure 23: **Fitted admixture model in the !Xuun.** See the caption to Figure 3 in the main text for details.

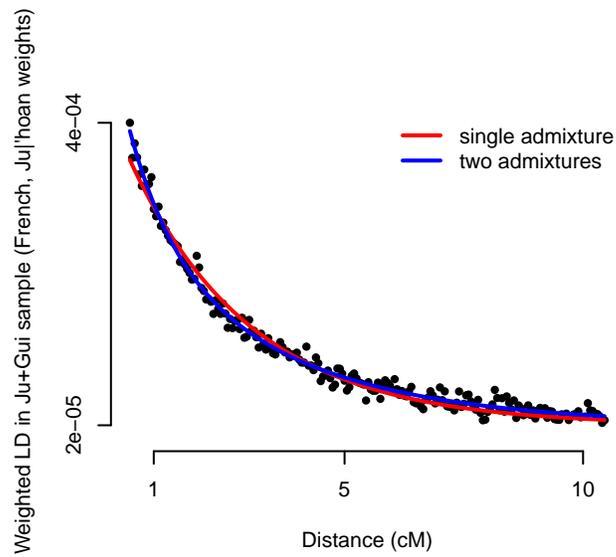
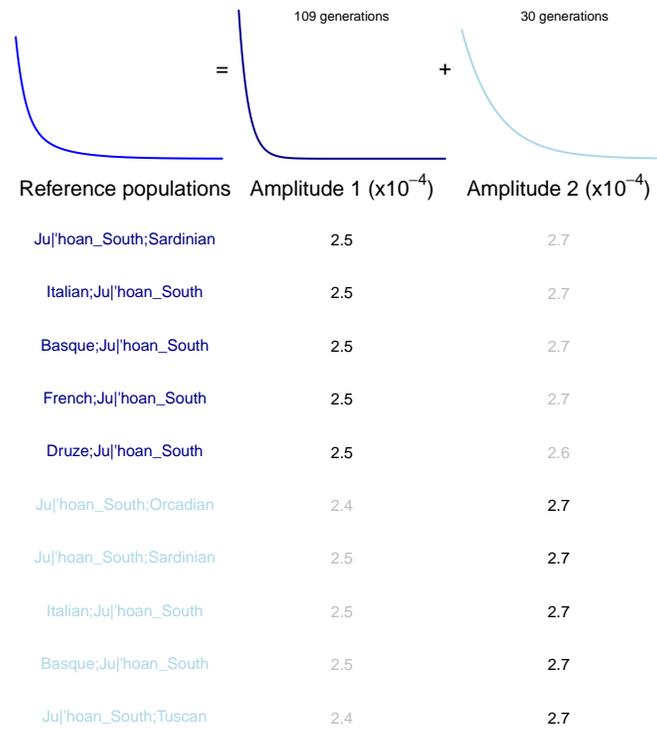

Figure 24: **Fitted admixture model in the combined Ju|'hoan_North and G|ui samples.** See the caption to Figure 3 in the main text for details.

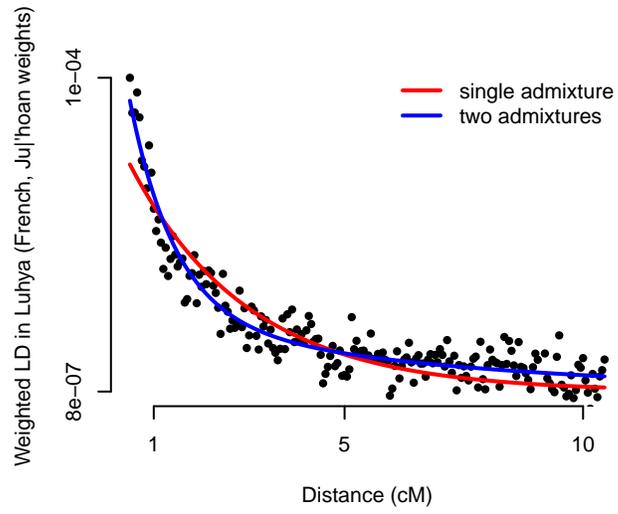

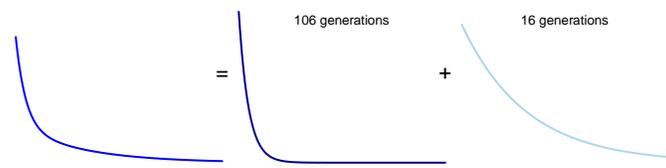

| | 106 generations | 16 generations |
|---|---|---|
| MbutiPygmy;Sardinian | 1.7 | 0.37 |
| Italian;MbutiPygmy | 1.6 | 0.37 |
| Basque;MbutiPygmy | 1.6 | 0.37 |
| Juhoansi;Sardinian | 1.5 | 0.33 |
| Cypriots;MbutiPygmy | 1.5 | 0.38 |
| BantuSouthAfrica;Cypriots | 1.3 | 0.45 |
| BantuSouthAfrica;Sardinian | 1.4 | 0.45 |
| BantuSouthAfrica;TSI | 1.3 | 0.45 |
| BantuSouthAfrica;Hungarians | 1.3 | 0.45 |
| Armenians;BantuSouthAfrica | 1.3 | 0.45 |

Figure 25: **Fitted admixture model in the Luhya.** See the caption to Figure 3 in the main text for details.

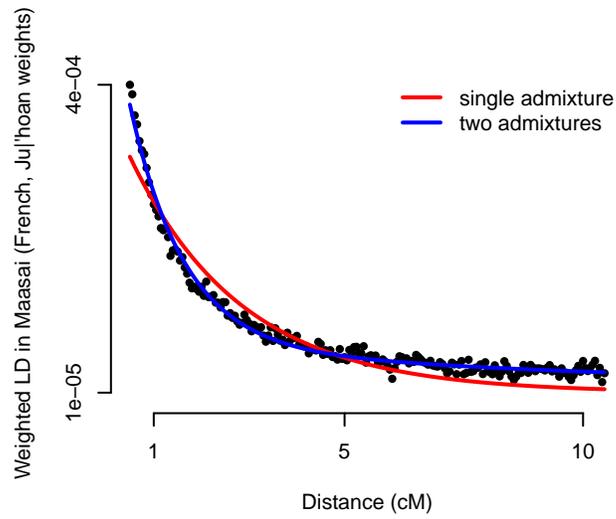

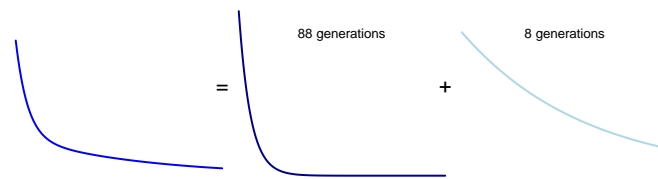

| | 88 generations | 8 generations |
|---|---|---|
| MbutiPygmy;Sardinian | 5.5 | 0.68 |
| Sardinian;SUDANESE | 5.4 | 0.56 |
| BiakaPygmy;Sardinian | 5.4 | 0.75 |
| ANUAK;Sardinian | 5.4 | 0.55 |
| BantuSouthAfrica;Sardinian | 5.4 | 0.74 |
| Basque;BiakaPygmy | 5.2 | 0.75 |
| BiakaPygmy;Italian | 5.2 | 0.75 |
| BiakaPygmy;TSI | 5.2 | 0.75 |
| BiakaPygmy;Sardinian | 5.4 | 0.75 |
| BiakaPygmy;Cypriots | 5.2 | 0.75 |

Figure 26: **Fitted admixture model in the Maasai.** See the caption to Figure 3 in the main text for details.

44

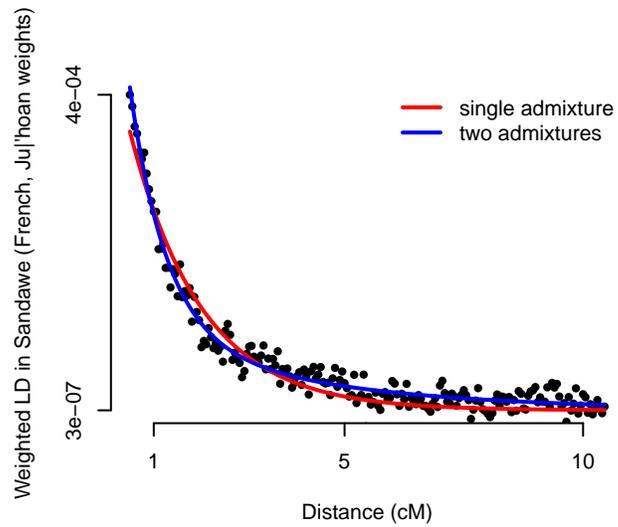

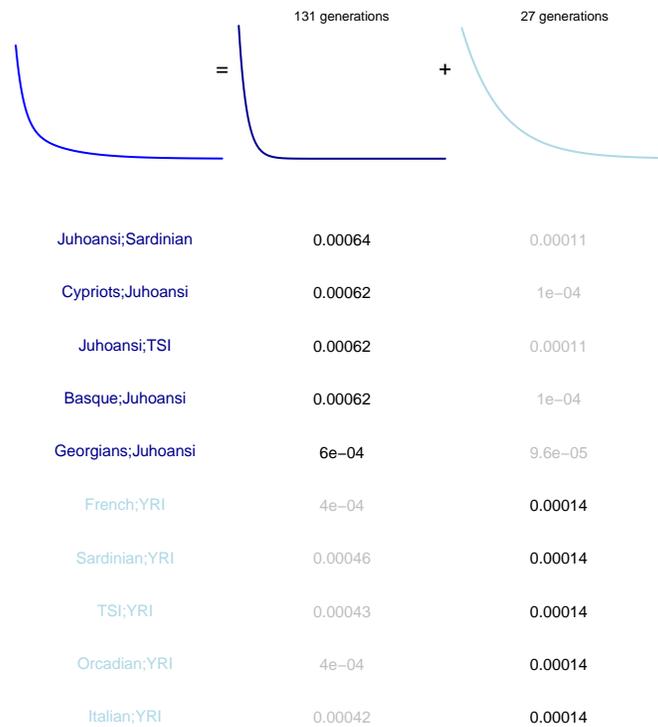

Figure 27: **Fitted admixture model in the Sandawe.** See the caption to Figure 3 in the main text for details.

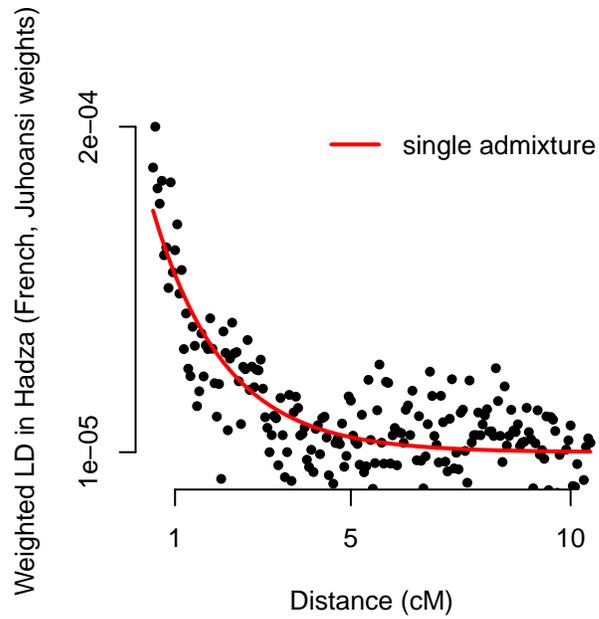

| Reference populations | Amplitude 1 (x$10^{-4}$) |
|---|---|
| MbutiPygmy;Sardinian | 1.9 |
| MbutiPygmy;TSI | 1.9 |
| Italian;MbutiPygmy | 1.9 |
| BiakaPygmy;Sardinian | 1.9 |
| Juhoansi;Sardinian | 1.9 |
| Basque;MbutiPygmy | 1.9 |
| French;MbutiPygmy | 1.9 |
| Juhoansi;TSI | 1.9 |
| BiakaPygmy;TSI | 1.9 |
| Hungarians;MbutiPygmy | 1.9 |

Figure 28: **Fitted admixture model in the Hadza.** See the caption to Figure 3 in the main text for details.

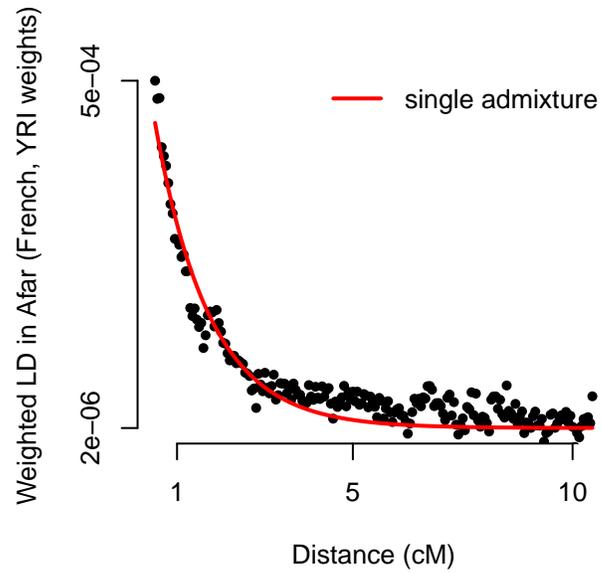

| Reference populations | Amplitude 1 (x$10^{-4}$) |
|---|---|
| Sardinian;SUDANESE | 7.9 |
| Italian;SUDANESE | 7.8 |
| SUDANESE;TSI | 7.8 |
| ANUAK;Sardinian | 7.7 |
| Orcadian;SUDANESE | 7.7 |
| French;SUDANESE | 7.7 |
| Cypriots;SUDANESE | 7.7 |
| ANUAK;Italian | 7.7 |
| ANUAK;TSI | 7.6 |
| CEU;SUDANESE | 7.6 |

Figure 29: **Fitted admixture model in the Afar.** See the caption to Figure 3 in the main text for details.

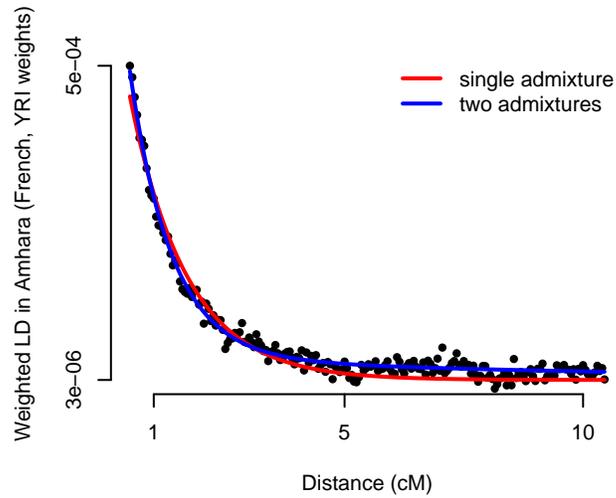

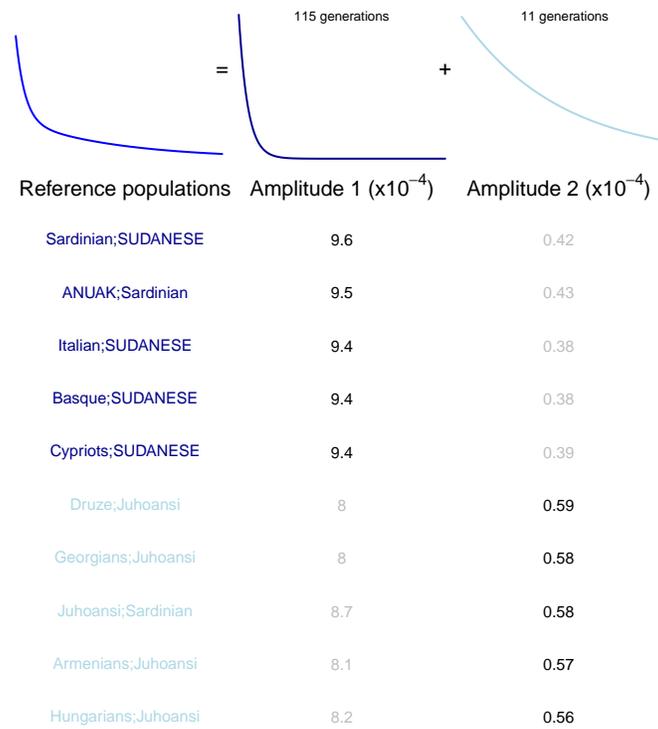

| Reference populations | Amplitude 1 (x10$^{-4}$) | Amplitude 2 (x10$^{-4}$) |
|---|---|---|
| Sardinian;SUDANESE | 9.6 | 0.42 |
| ANUAK;Sardinian | 9.5 | 0.43 |
| Italian;SUDANESE | 9.4 | 0.38 |
| Basque;SUDANESE | 9.4 | 0.38 |
| Cypriots;SUDANESE | 9.4 | 0.39 |
| Druze;Juhoansi | 8 | 0.59 |
| Georgians;Juhoansi | 8 | 0.58 |
| Juhoansi;Sardinian | 8.7 | 0.58 |
| Armenians;Juhoansi | 8.1 | 0.57 |
| Hungarians;Juhoansi | 8.2 | 0.56 |

Figure 30: **Fitted admixture model in the Amhara.** See the caption to Figure 3 in the main text for details.

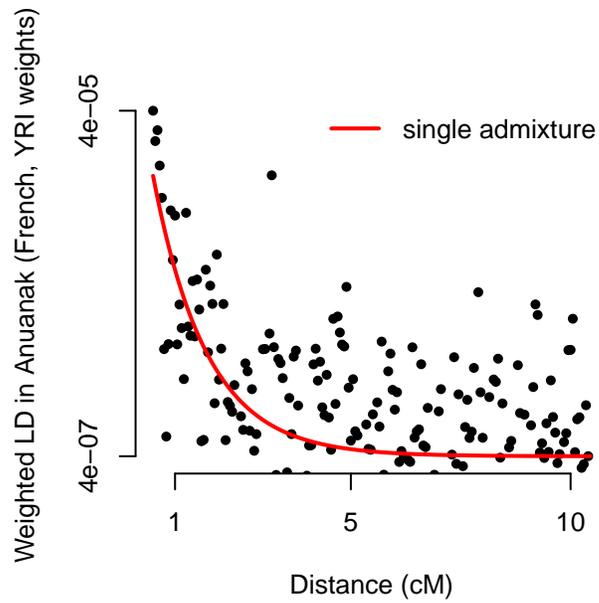

| Reference populations | Amplitude 1 (x$10^{-4}$) |
|---|---|
| Basque;SUDANESE | 0.67 |
| Italian;SUDANESE | 0.67 |
| French;SUDANESE | 0.66 |
| Hungarians;SUDANESE | 0.66 |
| SUDANESE;TSI | 0.65 |
| Russian;SUDANESE | 0.65 |
| CEU;SUDANESE | 0.65 |
| Orcadian;SUDANESE | 0.64 |
| Georgians;SUDANESE | 0.64 |
| Sardinian;SUDANESE | 0.63 |

Figure 31: **Fitted admixture model in the Anuak.** See the caption to Figure 3 in the main text for details.

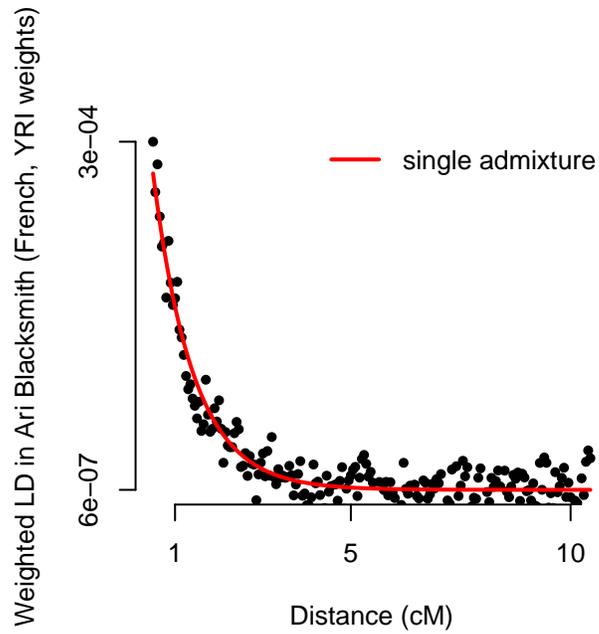

| Reference populations | Amplitude 1 (x$10^{-4}$) |
|---|---|
| MbutiPygmy;Sardinian | 5.2 |
| Juhoansi;Sardinian | 5.1 |
| Basque;MbutiPygmy | 5.1 |
| MbutiPygmy;TSI | 5 |
| Italian;MbutiPygmy | 5 |
| Cypriots;MbutiPygmy | 5 |
| Juhoansi;TSI | 5 |
| Basque;Juhoansi | 5 |
| BiakaPygmy;Sardinian | 5 |
| French;MbutiPygmy | 4.9 |

Figure 32: **Fitted admixture model in the Ari Blacksmiths.** See the caption to Figure 3 in the main text for details.

50

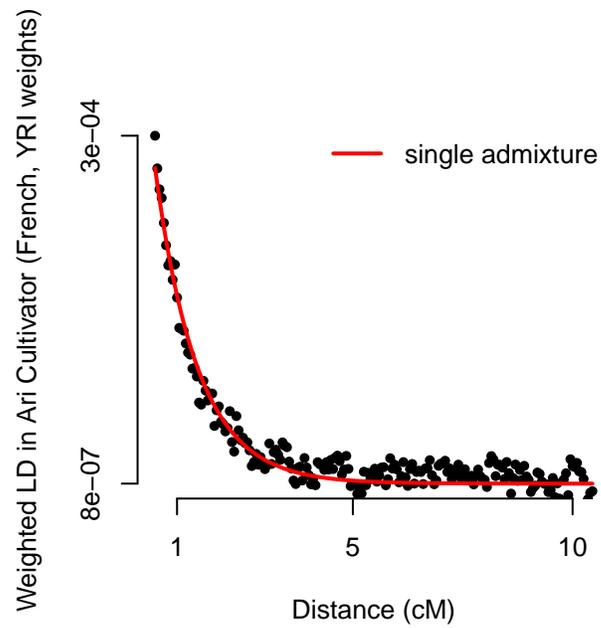

| Reference populations | Amplitude 1 (x$10^{-4}$) |
|---|---|
| Juhoansi;Sardinian | 5.6 |
| Juhoansi;TSI | 5.4 |
| Italian;Juhoansi | 5.4 |
| Cypriots;Juhoansi | 5.4 |
| MbutiPygmy;Sardinian | 5.4 |
| Basque;Juhoansi | 5.3 |
| Armenians;Juhoansi | 5.3 |
| Sardinian;Xun | 5.3 |
| Druze;Juhoansi | 5.3 |
| Hungarians;Juhoansi | 5.3 |

Figure 33: **Fitted admixture model in the Ari Cultivators.** See the caption to Figure 3 in the main text for details.

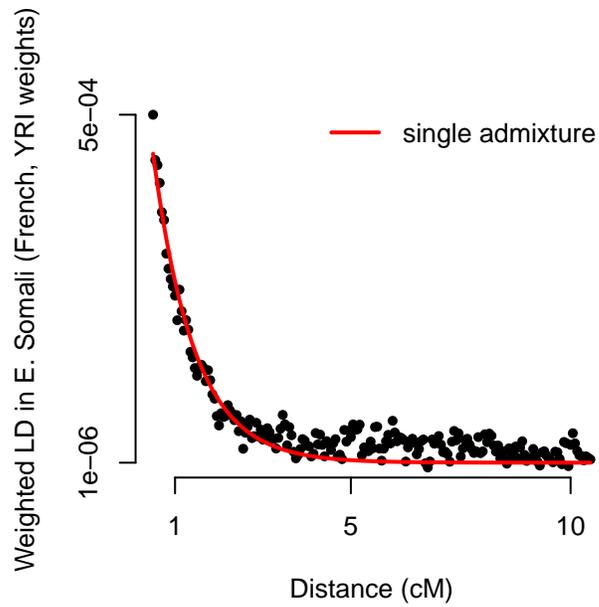

| Reference populations | Amplitude 1 (x$10^{-4}$) |
|---|---|
| ANUAK;Sardinian | 8 |
| Sardinian;SUDANESE | 8 |
| ANUAK;Italian | 7.9 |
| Italian;SUDANESE | 7.9 |
| ANUAK;TSI | 7.9 |
| SUDANESE;TSI | 7.8 |
| ANUAK;Basque | 7.8 |
| Basque;SUDANESE | 7.8 |
| BiakaPygmy;Sardinian | 7.8 |
| MbutiPygmy;Sardinian | 7.8 |

Figure 34: **Fitted admixture model in the Ethiopian Somali.** See the caption to Figure 3 in the main text for details.

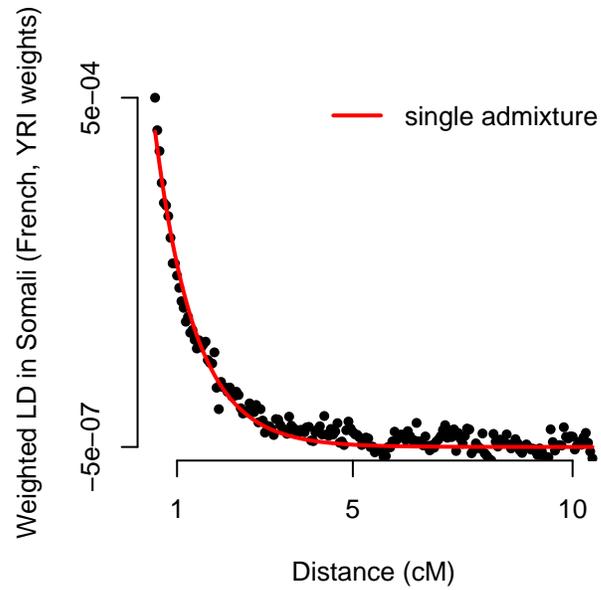

| Reference populations | Amplitude 1 (x$10^{-4}$) |
|---|---|
| Sardinian;SUDANESE | 8.6 |
| ANUAK;Sardinian | 8.5 |
| Italian;SUDANESE | 8.5 |
| SUDANESE;TSI | 8.4 |
| ANUAK;Italian | 8.4 |
| Cypriots;SUDANESE | 8.4 |
| Basque;SUDANESE | 8.4 |
| ANUAK;TSI | 8.3 |
| French;SUDANESE | 8.3 |
| MbutiPygmy;Sardinian | 8.3 |

Figure 35: **Fitted admixture model in the Somali.** See the caption to Figure 3 in the main text for details.

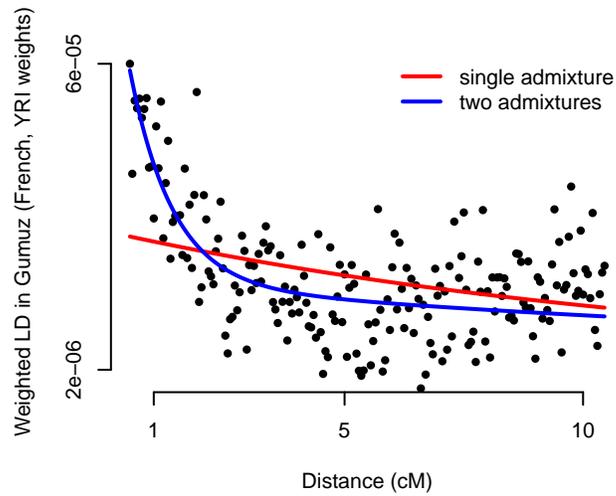

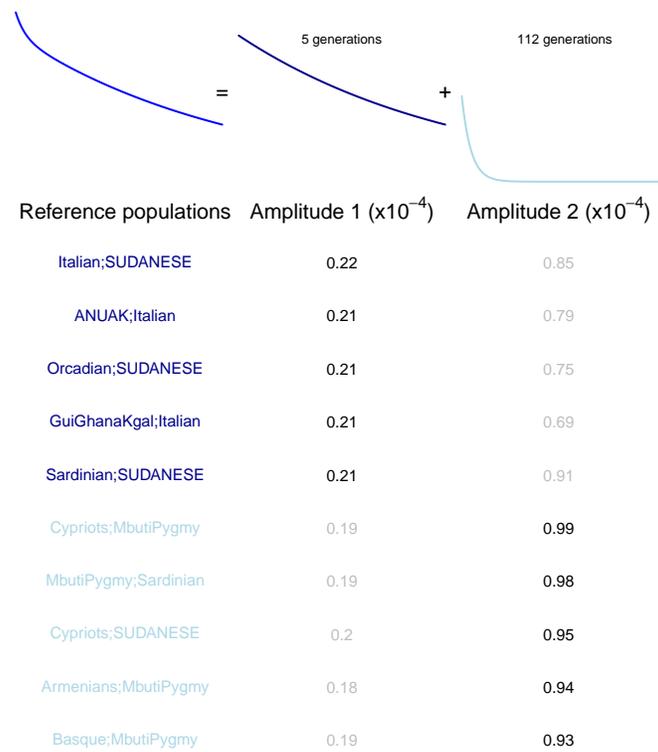

| Reference populations | Amplitude 1 (x$10^{-4}$) | Amplitude 2 (x$10^{-4}$) |
|---|---|---|
| Italian;SUDANESE | 0.22 | 0.85 |
| ANUAK;Italian | 0.21 | 0.79 |
| Orcadian;SUDANESE | 0.21 | 0.75 |
| GuiGhanaKgal;Italian | 0.21 | 0.69 |
| Sardinian;SUDANESE | 0.21 | 0.91 |
| Cypriots;MbutiPygmy | 0.19 | 0.99 |
| MbutiPygmy;Sardinian | 0.19 | 0.98 |
| Cypriots;SUDANESE | 0.2 | 0.95 |
| Armenians;MbutiPygmy | 0.18 | 0.94 |
| Basque;MbutiPygmy | 0.19 | 0.93 |

Figure 36: **Fitted admixture model in the Gumuz.** See the caption to Figure 3 in the main text for details.

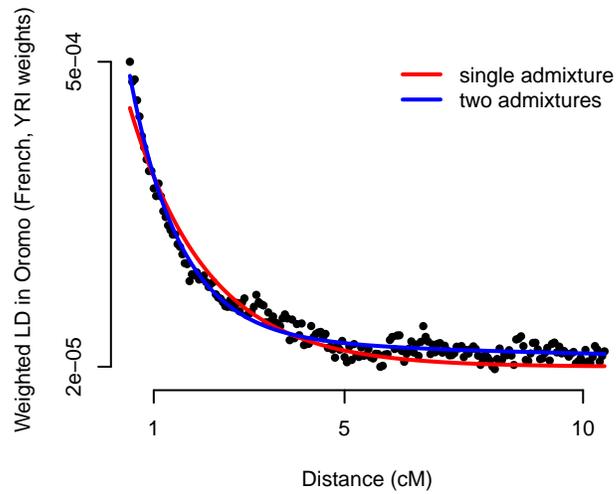
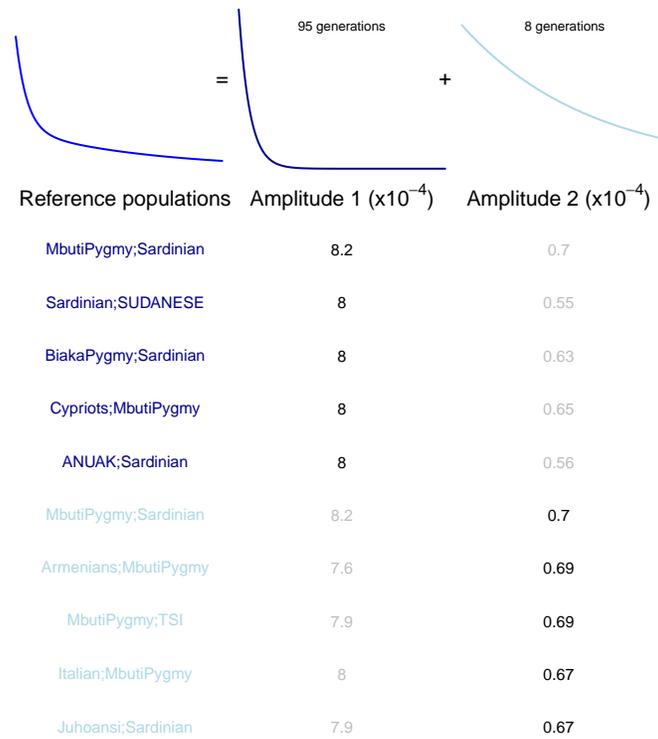

Figure 37: **Fitted admixture model in the Oromo.** See the caption to Figure 3 in the main text for details.

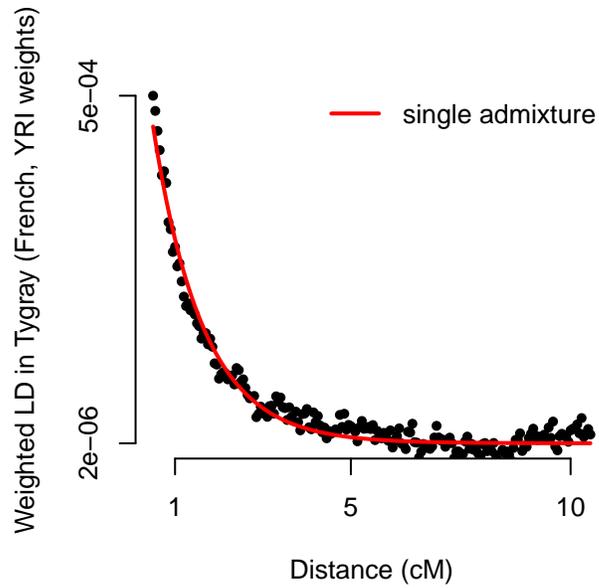

| Reference populations | Amplitude 1 (x$10^{-4}$) |
|---|---|
| Sardinian;SUDANESE | 8.7 |
| ANUAK;Sardinian | 8.7 |
| Cypriots;SUDANESE | 8.7 |
| ANUAK;Cypriots | 8.6 |
| MbutiPygmy;Sardinian | 8.6 |
| Italian;SUDANESE | 8.6 |
| ANUAK;Italian | 8.5 |
| SUDANESE;TSI | 8.5 |
| ANUAK;TSI | 8.5 |
| Cypriots;MbutiPygmy | 8.5 |

Figure 38: **Fitted admixture model in the Tygray.** See the caption to Figure 3 in the main text for details.

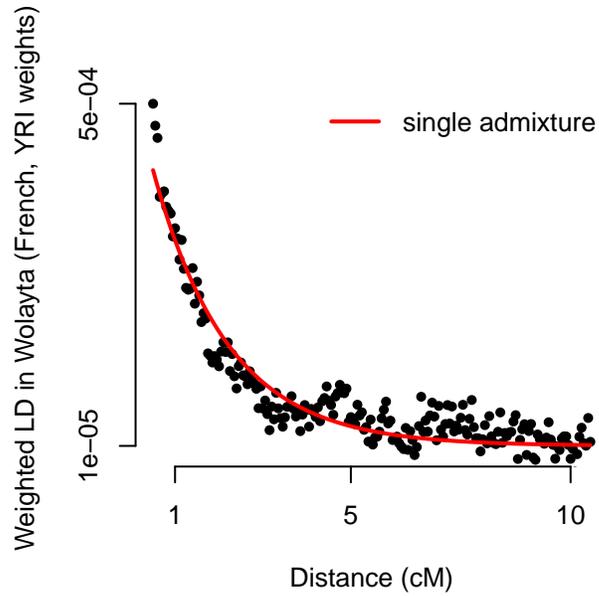

| Reference populations | Amplitude 1 (x$10^{-4}$) |
|---|---|
| Cypriots;Juhoansi | 6.5 |
| Juhoansi;Sardinian | 6.5 |
| MbutiPygmy;Sardinian | 6.5 |
| Italian;Juhoansi | 6.5 |
| Cypriots;MbutiPygmy | 6.5 |
| Italian;MbutiPygmy | 6.4 |
| Juhoansi;TSI | 6.4 |
| MbutiPygmy;TSI | 6.3 |
| Armenians;Juhoansi | 6.3 |
| French;Juhoansi | 6.3 |

Figure 39: **Fitted admixture model in the Wolayta.** See the caption to Figure 3 in the main text for details.

57

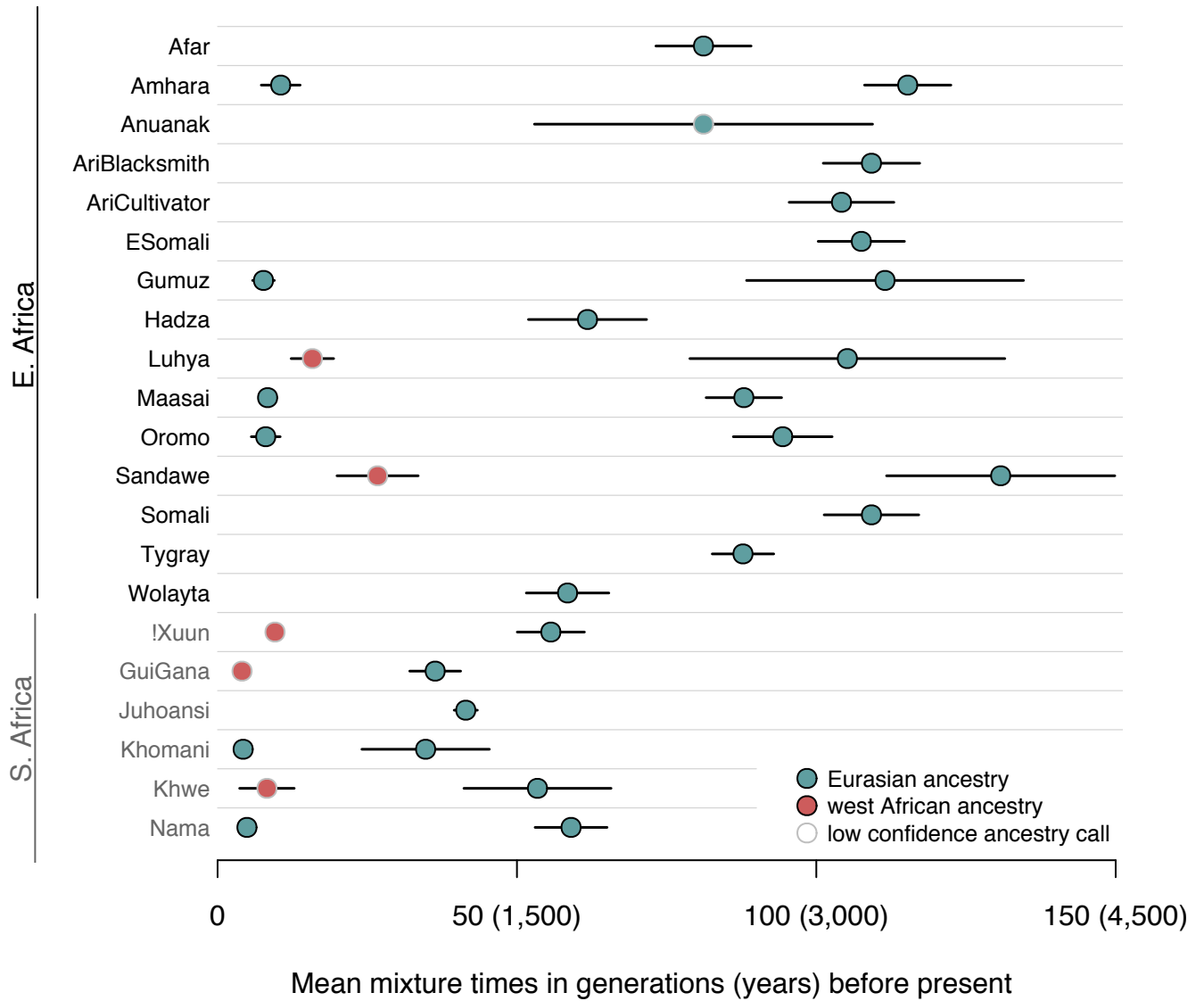

Figure 40: **Inferred times of admixture in southern and eastern Africa.** We applied our method to estimate the number of mixture events in the history of a population and their times; plotted are the estimated times. Lines represent a single standard error. The eastern African populations are the same as those in Figure 4 in the main text, while the southern African populations are those from Schlebusch et al. [2012]. In this figure the estimates in the southern and eastern African populations come from the exact same set of SNPs.

58

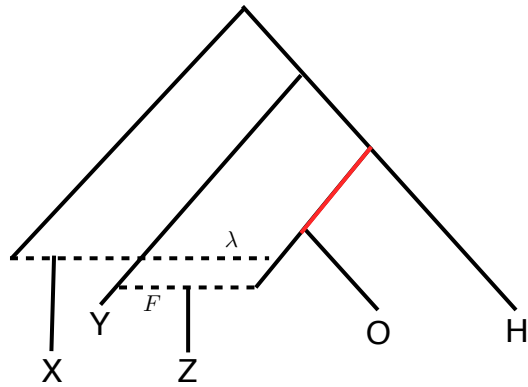

Figure 41: **Assumed population phylogeny for $f_4$ estimation of west Eurasian ancestry in African populations.** To calculate the proportion of west Eurasian ancestry in each southern and eastern African population, we used the following phylogeny for the Khoisan populations. $X$ represents the test population, $Y$ represents the Yoruba, $Z$ represents either the Sardinians or Druze, $O$ represents the Orcadians, $H$ represents the Han, $\lambda$ represents the proportions of west Eurasian ancestry in $X$, and $F$ represents the proportion of Yoruba-like ancestry in $Z$. The red branch is the relevant one for estimating west Eurasian ancestry. If we let $l$ be the length of the red branch, $f_4(H, O; X, Z) = (1 - \lambda - F)l$, and $f_4(H, O; Y, Z) = (1 - F)l$. Thus, the $f_4$ ratio $\frac{f_4(H,O;X,Z)}{f_4(H,O;Y,Z)} = \frac{1-\lambda-F}{1-F}$.


\clearpage
\end{document}